\documentclass[a4paper, oneside, 12pt]{article}%
\usepackage{amsmath}
\usepackage{amsfonts}
\usepackage{amssymb}
\usepackage{graphicx}
\usepackage{appendix}
\usepackage[square, numbers, sort&compress]{natbib}
\usepackage{color}%
\providecommand{\U}[1]{\protect\rule{.1in}{.1in}}

\newtheorem{theorem}{Theorem}[section]

\newtheorem{definition}[theorem]{Definition}

\newtheorem{lemma}[theorem]{Lemma}

\newtheorem{remark}[theorem]{Remark}

\numberwithin{equation}{section}

\newcommand{\E}{{\mathbb E}}
\newcommand{\R}{{\mathbb R}}
\newcommand{\pf}{\noindent\textbf{Proof:} }
\newcommand{\eof}{\hfill{$\Box$}}

\newcommand{\BMO}{L^{2, \;\mathrm{BMO}}_{\mathbb F}(0, T;\mathbb{R}^{n})}
\usepackage{url}
\usepackage{hyperref}
\hypersetup{
colorlinks=true, 
breaklinks=true, 
urlcolor=green , 
linkcolor=red, 
citecolor=blue, 
pdftitle={}, 
pdfauthor={}, 
pdfsubject={} 
}
\newcommand{\esssup}{\ensuremath{\operatorname*{ess\:sup}}}

\newcommand{\argmin}{\ensuremath{\operatorname*{argmin}}}

\newcommand{\ep}{\varepsilon}

\usepackage[normalem]{ulem}

\renewcommand{\geq}{\geqslant}
\renewcommand{\leq}{\leqslant}

\newcommand{\lnu}{\ensuremath{L^{2,\nu}}}

\newcommand{\ltwonu}{\ensuremath{L^{2, \nu}_{\mathcal{P}}(0, T;\mathbb{R})}}
\newcommand{\linnu}{\ensuremath{L^{\infty,\nu}_{\mathcal{P}}(0, T;\mathbb{R})}}
\newcommand{\nn}{\nonumber}

\allowdisplaybreaks[3] 

\begin{document}

\title{Constrained mean-variance investment-reinsurance under the Cram\'er-Lundberg model with random coefficients}
\author{Xiaomin Shi\thanks{School of Statistics and Mathematics, Shandong University of Finance and Economics, Jinan 250100, China.  Email: \textit{shixm@mail.sdu.edu.cn}}
\and Zuo Quan Xu\thanks{ 
Department of Applied Mathematics, The Hong Kong Polytechnic University, Kowloon, Hong Kong, China. Email: \textit{maxu@polyu.edu.hk}}
}
\maketitle

In this paper, we study an optimal mean-variance investment-reinsurance problem for an insurer (she) under a Cram\'er-Lundberg model with random coefficients. At any time, the insurer can purchase reinsurance or acquire new business and invest her surplus in a security market consisting of a risk-free asset and multiple risky assets, subject to a general convex cone investment constraint. We reduce the problem to a constrained stochastic linear-quadratic control problem with jumps whose solution is related to a system of partially coupled stochastic Riccati equations (SREs). Then we devote ourselves to establishing the existence and uniqueness of solutions to the SREs by pure backward stochastic differential equation (BSDE) techniques. We achieve this with the help of approximation procedure, comparison theorems for BSDEs with jumps, log transformation and BMO martingales. The efficient investment-reinsurance strategy and efficient mean-variance frontier are explicitly given through the solutions of the SREs, which are shown to be a linear feedback form of the wealth process and a half-line, respectively.

\smallskip
{\textbf{Keywords:}} Mean-variance investment-reinsurance; convex cone constraints; random coefficients; partially coupled stochastic Riccati equations; backward stochastic differential equations with jumps

\smallskip
\textbf{Mathematics Subject Classification (2020)} 93E20; 60H30; 91G10

\addcontentsline{toc}{section}{\hspace*{1.8em}Abstract}

\section{Introduction}

The use of reinsurance, in combination with investment, has become a popular strategy for insurers to manage their risk exposure in recent years. In the literature, there are three types of criteria that are widely used to determine the optimal investment-reinsurance strategies for insurers: expected utility maximization, minimizing the ruin probability, and the mean-variance criterion. As too many contributions have been made in this field, we will only mention few of them that are directly related to the mean-variance criterion, which is exactly the criterion adopted in this paper.

With a risk process described by the classical Cram\'er-Lundberg model, Bauerle \cite{Ba} solves the problem of finding the optimal reinsurance strategy which minimizes the expected quadratic distance of the risk reserve to a given benchmark. Bai and Zhang \cite{BZ} investigate an optimal investment-reinsurance problem under the mean-variance criterion. They derive the efficient strategy and efficient frontier explicitly via a verification theorem and viscosity solutions of Hamilton-Jacobi-Bellman (HJB) equations, taking the advantage of that the market coefficients are constants. Bi and Guo \cite{BG} examine the same problem and incorporate jumps in the stock price. With a risk process approximated by a Brownian motion with drift, Chen and Yam \cite{CY} consider this problem when the coefficients are switched according to a Markov chain.
Shen and Zeng \cite{SZ} study this problem in a stochastic factor model. Neither \cite{CY} nor \cite{SZ} has put constraints on the investment strategies.

In a typical financial market, the excess return rates and volatility rates of stocks may depend on their historical values, which renders the markets coefficients non-Markovian random processes. 
The aim of this paper is to extend the models of \cite{BZ} and \cite{BG}, which have deterministic coefficients, one risky asset, and no-shorting constraints, to include random coefficients (including excess return rates and volatility rates), multiple risky assets, and general convex cone investment constraints. Along this generalization, the HJB equation approach used in \cite{BZ} and \cite{BG} is not applicable anymore. One strong motivation of conducting this research comes from the statement \emph{``the financial market parameters... can change from deterministic functions to general stochastic processes. In this case, it is not easy to get the explicit expressions of the efficient frontier and the efficient strategy''} in \cite[Concluding Remarks]{BG} and the statement \emph{``due to positive reinsurance... and no-shorting restrictions. Thereby, the elegant Riccati equation...can not be applied''} in \cite[Page 182]{BZ}. We will show in addition that the Riccati equation method still works with general convex cone investment constraints. Our generalization allows for a more comprehensive exploration of the impact of uncertainty and market risk on the investment-reinsurance strategy of insurers.

We plan to tackle the problem via solving the famous stochastic Riccati equations (SREs). Even without investment constraints, the intrinsic constraint on the reinsurance policy renders the problem a constrained one. It is the randomness of the market coefficients that brings a jump dynamics into the SREs. This makes them extremely difficult to study.

SREs are intensively studied in the literature; see, e.g., Hu and Zhou \cite{HZ}, Kohlmann and Tang \cite{KT}, Sun, Xiong and Yong \cite{SXY}, Tang \cite{Tang}, to name a few. But no jump processes are involved in their SREs. Dong \cite{Dong}, Kharroubi, Lim and Ngoupeyou \cite{KLN} study SREs with a single jump, in the filtration enlargement theory, whose solution could be constructed via backward stochastic differential equations (BSDEs) driven only by Brownian motion (without jumps), but they do not provide any uniqueness results about the solutions of these equations. Zhang, Dong and Meng \cite{ZDM} make a great progress in studying matrix-valued SREs with jumps. Their results, however, do not cover our results, because their results require a uniform positive control weight, but in our problem the weight is 0. Also, neither \cite{KLN} nor \cite{ZDM} consider control constraints.

Our SREs consist of a pair of partially coupled BSDEs \eqref{P1} and \eqref{P2}. Please note that jump terms are presented in them. There is a quite significant body of literature on BSDEs with jumps and on their applications to portfolio optimization problems. To name a few, Tang and Li \cite{TL} firstly obtain the existence and uniqueness of solutions to BSDEs with jumps and Lipschitz generators, followed notably by Royer \cite{Royer}, Quenes and Sulem \cite{QS}, Yao \cite{Yao}, etc. BSDEs with jumps and quadratic generators and their applications in expected utility maximization problems have also been investigated; see, e.g., Antonelli and Mancini \cite{AM}, Kazi-Tani, Possamai and Zhou \cite{KPZ}, Laeven and Stadje \cite{LS}, Morlais \cite{Mo, Mo2}. Regarding our BSDEs with jumps, some parts of solution will appear in the denominator of the generators, thus they do not fit the usual quadratic growth condition and also violate the locally Lipschitz condition which are required in the aforementioned results. Therefore, no results in the literature, to the best of our knowledge, can cover our partially coupled BSDE with jumps. One of the main technical contributions of this paper is to establish the solvability (including the existence and uniqueness) of these new BSDEs. This is of great importance in its own interest from the BSDE theory point of view. Besides the standard truncation and approximation techniques, BMO martingale theory, a critical tool used to tackle the SREs is Quenes and Sulem's \cite{QS} comparison theorem for BSDEs with jumps. Although Royer \cite{Royer} proves a comparison theorem BSDEs with jumps, his result requires that $\rho$ in \eqref{comparisonkey} has a uniform lower bound $>-1$ in order to apply Girsanov's theorem. But the requirement is just failed to be satisfied by our $\rho$ (which is merely $\geq -1$). It is Quenes and Sulem's comparison theorem, which releases the lower bound to be $\geq -1$, allows us to tackle the problem.

After solving the SREs, we construct a feedback control based on their solutions, which is shown to be optimal for the original constrained stochastic linear-quadratic control problem with jumps by a verification theorem. Furthermore, we are able to present the efficient investment-reinsurance strategy and the efficient frontier in closed forms for the original mean-variance problem.

The rest of this paper is organized as follows. In Section \ref{fm}, we present the financial market and formulate a constrained mean-variance investment-reinsurance problem. In Section \ref{re}, we introduce and solve a relaxed problem by solving the related SREs. In Section \ref{solution}, we present the efficient investment-reinsurance strategy and efficient frontier for the original mean-variance problem. We conclude the paper in Section \ref{cr}. Appendix \ref{appnA} collects some definitions and properties of continuous BMO martingales. A lengthy proof of Theorem \ref{Th:P1} is relegated to Appendix \ref{appnB}.

\section{Problem formulation}\label{fm}
Let $(\Omega, \mathcal F, \mathbb{P})$ be a fixed complete probability space on which is defined a standard $n$-dimensional Brownian motion $W$. We denote by $\R^m$ the set of $m$-dimensional column real vectors, by $\R^m_+$ (resp. $\R^m_-$) the set of vectors in $\R^m$ whose components are nonnegative (resp. nonpositive), by $\R^{m\times n}$ the set of $m\times n$ real matrices, and by $\mathbb{S}^n$ the set of symmetric $n\times n$ real matrices. For $M=(m_{ij})\in \R^{m\times n}$, we denote its transpose by $M^{\top}$, and its norm by $|M|=\sqrt{\sum_{ij}m_{ij}^2}$. If $M\in\mathbb{S}^n$ is positive definite (resp. positive semidefinite), we write $M>$ (resp. $\geq$) $0.$ We write $A>$ (resp. $\geq$) $B$ if $A, B\in\mathbb{S}^n$ and $A-B>$ (resp. $\geq$) $0.$ As usual, we write $x^+=\max\{x, 0\}$ and $x^-=\max\{-x, 0\}$ for $x\in\R$.

Let
$\{N_t\}$ be a homogeneous Poisson process with intensity $\lambda>0$ which counts the number of claims. Let $Y_i$ denote the $i^{\mathrm{th}}$ claim, $i=1,2,...$, and assume they are bounded\footnote{Technically, we need this condition so that the comparison theorem for BSDE with jumps (see, e.g., \cite[Theorem 4.2]{QS}) can be applied to prove Theorem \ref{Th:P2}. From insurance practice point of view, an insurance claim shall never exceed the value of the insured asset, which is usually upper bounded by some constant (such as the value of a new replacement of the insured car or house); even if the insured asset (such as human life) is invaluable, there is usually a maximum payment amount in insurance contract. Hence, it is reasonable to assume that the claims are bounded.}
independent and identically distributed nonnegative random variables with a common probability distribution function $\nu:\R_+\rightarrow[0,1]$. The first and second moments of the claims are denoted by
\begin{align}\label{def:bY}
b_Y=\int_{\R_+} y \nu(dy)>0,\quad \sigma_Y^2=\int_{\R_+} y^2 \nu(dy)>0.
\end{align}
In the classical Cram\'er-Lundberg model, the surplus $R_t$ dynamic without reinsurance or investment follows
\begin{align}\label{Rprocess}
R_t=R_0+pt-\sum_{i=1}^{N_t} Y_i, 
\end{align}
where $p$ is the premium rate which is assumed to be calculated according to the expected value principle, i.e., $$p=(1+\eta)\lambda b_Y>0$$ and $\eta>0$ is the relative safety loading of the insurer. As in \cite{SZ}, we use a Poisson random measure $\gamma$ on $\Omega\times[0,T]\times\R_+$ to denote the compound Poisson process $\sum_{i=1}^{N_t} Y_i$ as
\begin{align*}
\int_0^t\int_{\R_+}y\gamma(ds,dy)=\sum_{i=1}^{N_t} Y_i, \ \forall\; t\in[0,T].
\end{align*}
Assume that $N, Y_i, i=1,2,...$ are independent, then
\begin{align*}
\int_0^t\int_{\R_+}y\lambda\nu(dy)=\E\Big[\sum_{i=1}^{N_t} Y_i\Big], \ \forall\; t\in[0,T].
\end{align*}
The compensated Poisson random measure is denoted by $$\tilde \gamma(dt,dy)=\gamma(dt,dy)-\lambda\nu(dy)dt.$$

We suppose that the Brownian motion $W$ and the Poisson random measure $\gamma(dt,dz)$ are independent. Define the filtration $\mathbb{F}=\{\mathcal{F}_t,t\geq0\}$ as the augmented natural filtration associate with $W$ and $\gamma$.

Let $\mathcal{P}$ be the $\mathbb{F}$-predictable $\sigma$-field on $[0,T]\times\Omega$, $\mathcal{B}(\mathbb{R_+})$ the Borel $\sigma$-algebra of $\mathbb{R_+}$. We use the following spaces:
\begin{align*}
L^{2}_{\mathbb F}(0, T;\mathbb{R})=\Big\{\phi &: [0, T]\times\Omega\rightarrow
\mathbb{R}\;\Big|\;\phi\mbox{ is $\mathcal{P}$-measurable and } \E\Big[\int_{0}^{T}|\phi_t|^{2}dt\Big]<\infty\Big\}, \\
\lnu=\Big\{\phi &:\R_+\rightarrow\mathbb{R}\;\Big|\;\phi\mbox{ is $\mathcal{B}(\R_+)$-measurable and $\int_{\R_+} |\phi(y)|(1+y^2)\nu(dy)<\infty$}\Big\},\\
L^{2,\nu}_{\mathcal{P}}(0, T;\mathbb{R})=\Big\{\phi &:[0, T]\times\Omega\times \R_+\rightarrow
\mathbb{R}\;\Big|\; \phi\mbox{ is $\mathcal{P}\otimes\mathcal{B}(\R_+)$-measurable }\\
&\quad\mbox{and } \E\Big[\int_{0}^{T}\int_{\R_+}|\phi_t(y)|^{2}\lambda\nu(dy)dt\Big]<\infty\Big\},\\
\linnu =\Big\{\phi &\in L^{2,\nu}_{\mathcal{P}}(0, T;\mathbb{R})\Big|\mbox{ $\phi$ is essentially bounded w.r.t.} \ dt\otimes d\mathbb{P}\otimes d\nu\Big\},\\
S^{2}_{\mathbb{F}}(0,T;\mathbb{R})=\Big\{\phi &:[0,T]\times\Omega
\rightarrow\mathbb{R}\;\Big|\;(\phi_{t})_{0\leq t\leq T}\mbox{is c\`ad-l\`ag $\mathbb{F}$-adapted}\\
&\quad\mbox{and} \ \E\Big[\sup_{0\leq t\leq T}|\phi_t|^2\Big]<\infty\Big\},\\
S^{\infty}_{\mathbb{F}}(0,T;\mathbb{R})=\Big\{\phi &\in S^{2}_{\mathbb{F}}(0,T;\mathbb{R})\Big| \mbox{ $\phi$ is essentially bounded w.r.t.} \ dt\otimes d\mathbb{P}\Big\}, \\
\BMO=\bigg\{\Lambda &\in L^{2}_{\mathbb F}(0, T;\mathbb{R}^{n}) \;\bigg|\; \int_0^\cdot\Lambda_s^{\top}dW_s \mbox{ is a BMO martingale on $[0, T]$}\bigg\}.
\end{align*}
We recall some basic facts about continuous BMO martingales that will be used in our analysis in Appendix \ref{appnA}. Please refer to Kazamaki \cite{Ka} for a systematic account on continuous BMO martingales and He, Wang and Yan \cite{HWY} on c\`adl\`ag ones. The above definitions are generalized in the obvious way to the cases that $\mathbb{R}$ is replaced by $\mathbb{R}^n$, $\mathbb{R}^{n\times m}$ or $\mathbb{S}^n$. In our argument, $s$, $t$, $\omega$, ``almost surely'' and ``almost everywhere'', will be suppressed for simplicity in many circumstances, when no confusion occurs. If we say an inequality or equation that involves some function in $\ltwonu$ holds, then we mean it holds $dt\otimes d\mathbb{P}\otimes d\nu$ almost surely. Similarly, we say a process is bounded if it is essentially bounded with respect to (w.r.t.) $dt\otimes d\mathbb{P}\otimes d\nu$.

We consider an insurer (she) who is allowed to purchase proportional reinsurance or acquire new business to control its exposure to the insurance risk. Let $q_t$ be the value of risk exposure, which represents the insurer's retention level of insurance risk at time $t$. When $q_t\in[0,1]$, it corresponds to a proportional reinsurance cover; in this case, the insurer only need pays $100 q_t\%$ of each claim, and the reinsurer pays the rest; as a remedy, the insurer diverts part of the premium to the reinsurer at the rate of $(1-q_t)(1+\eta_r)\lambda b_Y,$ where $\eta_r\geq\eta$ is the reinsurer's relative security loading. When $q_t>1$, it corresponds to acquiring new business. The process $q_t$ is called a reinsurance strategy for convenience. By adopting a reinsurance strategy $q_t$, the insurer's surplus process $R^q$ follows
\begin{align*}
dR^q_t&=(1+\eta)\lambda b_{Y}dt-(1-q_t)(1+\eta_r)\lambda b_{Y}dt-q_t\int_{\R_+}y\gamma(dt,dy)\\
&=(\eta_r q_t+\eta-\eta_r)\lambda b_Ydt-q_t\int_{\R_+}y\tilde \gamma(dt,dy).
\end{align*}

The insurer is allowed to invest her wealth in a financial market consisting of a risk-free asset (the money market instrument or bond) whose price is $S_{0}$ and $m$ risky securities (the stocks) whose prices are $S_{1}, \ldots, S_{m}$. And assume $m\leq n$, i.e., the number of risky securities is no more than the dimension of the Brownian motion. Hence, the market is incomplete. The asset prices $S_k$, $k=0, 1, \ldots, m, $ are driven by SDEs:
\begin{align*}
\begin{cases}
dS_{0,t}=r_tS_{0,t}dt, \\
S_{0,0}=s_0,
\end{cases}
\end{align*}
and
\begin{align*}
\begin{cases}
dS_{k,t}=S_{k,t}\Big((\mu_{k,t}+r_t)dt+\sum\limits_{j=1}^n\sigma_{kj,t}dW_{j,t}\Big), \\
S_{k,0}=s_k,
\end{cases}
\end{align*}
where, for every $k=1, \ldots, m$, $r$ is the interest rate process, $\mu_k$ and $\sigma_k:=(\sigma_{k1}, \ldots, \sigma_{kn})$ are the mean excess return rate process and volatility rate process of the $k$th risky security. Define the mean excess return vector
$\mu=(\mu_1, \ldots, \mu_m)^{\top}$,
and the volatility matrix $\sigma= (\sigma_{kj})_{m\times n}.$
We assume, through this paper, the interest rate $r$ is a bounded deterministic measurable function of $t$, $\mu\in L_{\mathbb{F}}^\infty(0, T;\mathbb{R}^m)$ and
$\sigma\in L_{\mathbb{F}}^\infty(0, T;\mathbb{R}^{m\times n}).$ Furthermore, there exists a constant $\delta>0$ such that $\sigma\sigma^{\top}\geq \delta I_m$ for a.e. $t\in[0, T]$, where $I_m$ denotes the $m$-dimensional identity matrix.

The insurer's actions cannot affect the asset prices. She can decide at every time
$t\in[0, T]$ what amount $\pi_{j,t}$ of her wealth to invest in the $j$th risky asset, $j=1, \ldots, m$. The vector process $\pi:=(\pi_1, \ldots, \pi_m)^{\top}$ is called a portfolio of the investor. Then the investor's self-financing wealth process $X$ corresponding to an investment-reinsurance strategy $(\pi,q)$ satisfies the SDE:
\begin{align}
\label{wealth}
\begin{cases}
dX_t=[r_tX_{t-}+\pi_t^{\top}\mu_t+bq_t+a]dt+\pi_t^{\top}\sigma_tdW_t-q_t\int_{\R_+}y\tilde \gamma(dt,dy), \\
\; X_0=x,
\end{cases}
\end{align}
where
\begin{align}\label{def:ba}
b:=\lambda b_Y\eta_r>0,\quad a:=\lambda b_Y(\eta-\eta_r).
\end{align}
Let $\Pi$ be a given closed convex cone in $\mathbb{R}^m$. It is the constraint set for investment. The class of admissible investment-reinsurance strategies is defined as
\begin{align*}
\mathcal{U}:=\Big\{(\pi,q)\in L^2_\mathbb{F}(0, T;\mathbb{R}^{m+1})\;\Big|\; \pi \in\Pi, \ q\geq0, \mbox{ a.e. a.s.}\Big\}.
\end{align*}
For any admissible strategy $(\pi,q)\in\mathcal{U}$, the wealth process \eqref{wealth} admits a unique strong solution $X$, which will be denoted by $X^{\pi,q}$ whenever it is necessary to indicate its dependence on $(\pi,q)\in\mathcal{U}$.

For a given level $z\geq xe^{\int_0^Tr_sds}+a\int_0^Te^{\int_t^Tr_sds}dt,$
the insurer's investment-reinsurance mean-variance problem is to
\begin{align}
\mathrm{Minimize}&\quad \mathrm{Var}(X_T)\equiv\E\big[(X_T-\E(X_T))^{2}\big] =\E\big[X_T^{2}\big]-z^2 \label{optm}\\
\mathrm{ s.t.} &\quad
\begin{cases}
\E (X_T)=z, \\
(\pi,q)\in \mathcal{U}.\nn
\end{cases}
\end{align}
An optimal strategy $(\hat\pi,\hat q)$ to \eqref{optm}
is called an efficient/optimal strategy corresponding to $z$. And $\left(\sqrt{\mathrm{Var}(X_{T}^{\hat\pi,\hat q})},\ z\right)$ is
called an efficient mean-variance point. The set of all efficient mean-variance points corresponding to various rational expectation levels $z$:
$$\left\{\left(\sqrt{\mathrm{Var}(X_{T}^{\hat\pi,\hat q})},\ z\right)\;\bigg|\; z\geq xe^{\int_0^Tr_sds}+a \int_0^Te^{\int_t^Tr_sds}dt\right\}$$ is
called the efficient mean-variance frontier. The main goal of this paper is to determine the optimal investment-reinsurance strategy and the efficient frontier for the problem \eqref{optm}.

\section{Relaxed problem and its solution}\label{re}
To deal with the constraint $\E(X_T)=z$ in the problem \eqref{optm}, we introduce a Lagrange
multiplier $-2\zeta\in\mathbb{R}$ and obtain the following \emph{relaxed}
optimization problem:
\begin{align}\label{optmun}
\mathrm{Minimize} &\quad\; {\mathbb{E}}[X_T^2-z^2-2\zeta(X_T-z)]={\mathbb{E}}[(X_T-\zeta)^{2}]-(\zeta-z)^{2}=:\hat{J}(\pi, q, \zeta), \\
\mathrm{s.t.} &\quad (\pi,q)\in \mathcal{U}.\nn 
\end{align}
The associated value function is defined as
\begin{align}\label{def:J}
J(\zeta):=\inf_{(\pi,q)\in\mathcal{U}}\hat{J}(\pi, q,\zeta).
\end{align}
Because the mean-variance problem \eqref{optm} is a convex optimization problem, it is linked to the relaxed problem \eqref{optmun} by the Lagrange duality theorem (see Luenberger \cite{Lu})
\begin{align}\label{duality}
\min_{(\pi,q)\in\mathcal{U}, \E(X_T)=z}\mathrm{Var}(X_T)
&=\max_{\zeta\in\mathbb{R}}\min_{(\pi,q)\in\mathcal{U}}\hat{J}(\pi, q, \zeta)=\max_{\zeta\in\mathbb{R}}J(\zeta).
\end{align}
This allows us to solve the problem \eqref{optm} by a two-step procedure: First solving the relaxed problem \eqref{optmun} for any $\zeta\in\R$, then finding a $\zeta^{*}$ to maximize $J(\zeta)$.

\subsection{The SREs} 

We now focus on the relaxed problem \eqref{optmun}. This is a stochastic LQ problem with cone constraint. We plan to solve it via the famous SREs.

First for $(t,\omega,v,u,P_{1},\Lambda_{1},\Gamma_1,P_{2},\Lambda_{2},\Gamma_2)\in[0,T]\times\Omega\times\Pi\times\R_+\times
\R_+\times\R^m\times \lnu\times\R_+\times\R^m\times \lnu$, we define the following mappings:
\begin{align*}
F_1(t,\omega,v,P_1,\Lambda_1)&=P_1|\sigma^{\top}_tv|^2+2v^{\top}(P_1\mu_t+\sigma_t\Lambda_1),\\
F_2(t,\omega,v,P_2,\Lambda_2)&=P_2|\sigma^{\top}_tv|^2-2v^{\top}(P_2\mu_t+\sigma_t\Lambda_2),\\
G_1(u,P_1,\Gamma_1,P_2,\Gamma_2)&=\int_{\R_+}\Big[(P_1+\Gamma_1(y))\big[[(1-uy)^+]^2-1\big]+(P_2+\Gamma_2(y))[(1-uy)^-]^2\Big]\lambda \nu(dy)\\
&\quad+2uP_1(b+\lambda b_Y),
\end{align*}
and
\begin{align*}
F_1^*(t,\omega,P_1,\Lambda_1)&=\inf_{v\in\Pi}F_1(t,\omega,v,P_1,\Lambda_1),\\
F_2^*(t,\omega,P_2,\Lambda_2)&=\inf_{v\in\Pi}F_2(t,\omega,v,P_2,\Lambda_2),\\
G_1^*(P_1,\Gamma_1,P_2,\Gamma_2)&=\inf_{u\geq0}G_1(u,P_1,\Gamma_1,P_2,\Gamma_2), \\
G_2^*(P_2,\Gamma_2)
&=-\frac{\Big[\Big(P_2 b -\int_{\R_+}\Gamma_2(y) y\lambda\nu(dy)\Big)^+\Big]^2}
{\int_{\R_+} (P_2+\Gamma_2(y))y^2\lambda\nu(dy)}.
\end{align*}
If $P_1,P_2>0$, since $\sigma\sigma^{\top}\geq \delta I_m$, both $F_1^*(t,\omega,P_1,\Lambda_1)$ and $F_2^*(t,\omega,P_2,\Lambda_2)$ take finite values.
If $P_1+\Gamma_1, P_2+\Gamma_2>0$, then
\begin{align*}
G^*_1(P_1,\Gamma_1, P_2,\Gamma_2)\leq G_1(0,P_1,\Gamma_1,P_2,\Gamma_2)=0,
\end{align*}
and
\begin{align*}
G^*_1(P_1,\Gamma_1,P_2,\Gamma_2)
&\geq -\int_{\R_+}(P_1+\Gamma_1)\lambda \nu(dy)+2uP_1(b+\lambda b_Y).
\end{align*}
So $ G_1^*(P_1,\Gamma_1,P_2,\Gamma_2)$ also takes a finite value. Therefore, we can introduce the following SREs, a system of BSDEs, for our relaxed problem \eqref{optmun}:
\begin{align}
\label{P1}
\begin{cases}
dP_1=-\Big[2rP_1+F_1^*(t,P_1,\Lambda_{1})+G_1^*(P_1,\Gamma_1,P_2,\Gamma_2)\Big]dt+\Lambda_{1}^{\top}dW+\int_{\R_+}\Gamma_1(y)\tilde \gamma(dt,dy), \\
P_{1,T}=1, \\
P_{1,t}>0, \ P_{1,t-}+\Gamma_{1,t}>0, \ P_{2,t-}+\Gamma_{2,t}>0,\
\mbox{ for all $t\in[0, T]$,\qquad}
\end{cases}
\end{align}
and
\begin{align}
\label{P2}
\begin{cases}
dP_2=-\Big[2rP_2+F_2^*(t,P_2,\Lambda_{2})+G_2^*(P_2,\Gamma_2)\Big]dt+\Lambda_{2}^{\top}dW+\int_{\R_+}\Gamma_2(y)\tilde \gamma(dt,dy), \\
P_{2,T}=1, \\
P_{2,t}>0, \ \ P_{2,t-}+\Gamma_{2,t}>0, \ \mbox{ for all $t\in[0, T]$.}
\end{cases}
\end{align}

The remainder of this subsection will be devoted to resolving the solvability issues of the above SREs \eqref{P1} and \eqref{P2}. They are partially coupled in the sense that the former depends on the latter, but not vice versa. This partially coupling phenomenon is caused by the special kind of jumps, that is downward jumps, in the wealth dynamics \eqref{wealth}. With this kind of jumps, starting from a positive wealth level, one will eventually become negative over a sufficiently large time horizon, but from a negative wealth level, one will remain to be negative all the time.

We first give the precise definition of a solution to \eqref{P1} and \eqref{P2}.
\begin{definition}\label{def}
A stochastic process $(P_1,\Lambda_{1},\Gamma_1,P_2,\Lambda_{2},\Gamma_2)$ is called a solution to the system of BSDEs \eqref{P1} and \eqref{P2}, if it satisfies \eqref{P1} and \eqref{P2}, and $(P_1,\Lambda_{1},\Gamma_1,P_2,\Lambda_{2},\Gamma_2)\in S^{\infty}_{\mathbb{F}}(0,T;\mathbb{R})\times L^{2}_{\mathbb{F}}(0,T;\mathbb{R}^m)\times \linnu \times S^{\infty}_{\mathbb{F}}(0,T;\mathbb{R})\times L^{2}_{\mathbb{F}}(0,T;\mathbb{R}^m)\times \linnu $. A solution is called uniformly positive if $$P_{1,t},P_{1,t-}+\Gamma_{1,t},P_{2,t},P_{2,t-}+\Gamma_{2,t}\geq c,\quad\mbox{a.s.},$$ for all $t\in[0, T]$ with some positive constant $c$.\footnote{Hereafter, we shall use $c$ to represent a generic positive constant independent of $t$ and $\omega$, which can be different from line to line.}
\end{definition}

We will study the two equations \eqref{P1} and \eqref{P2} in a similar way. But \eqref{P1} is more complicated since it depends on the components of \eqref{P2}. Therefore, we first study \eqref{P1} in details, and then point out the major difference in solving \eqref{P2}.

\begin{theorem}\label{Th:P1}
Given a uniformly positive solution $(P_2,\Lambda_2,\Gamma_2)\in S^{\infty}_{\mathbb{F}}(0, T;\mathbb{R})\times L^{2}_{\mathbb{F}}(0, T;\mathbb{R}^n)\times \linnu $ to \eqref{P2}, the BSDE \eqref{P1} admits a unique uniformly positive solution $(P_1,\Lambda_1,\Gamma_1)\in S^{\infty}_{\mathbb{F}}(0, T;\mathbb{R})\times L^{2}_{\mathbb{F}}(0, T;\mathbb{R}^n)\times \linnu $.
\end{theorem}
\pf 
For integers $k=1,2,...$, define 
\begin{align*}
F_1^{*,k}(t,\omega,P_{1},\Lambda_{1})&=\inf_{v\in\Pi,|v|\leq k}
F_1(t,\omega,v,P_{1},\Lambda_{1}),\\
G_1^{*,k}(P_1,\Gamma_1,P_2,\Gamma_2)&=\inf_{0\leq u\leq k} G_1(u,P_1,\Gamma_1,P_2,\Gamma_2).
\end{align*}
Obviously, they are decreasing to $F_1^{*}(t,\omega,P_{1},\Lambda_{1})$ and $G_1^{*}(P_1,\Gamma_1,P_2,\Gamma_2)$, respectively, as $k$ goes to infinite, when $P_1,P_1+\Gamma_1>0$. By virtue of the truncation effect of $k$ and the boundedness of $(P_2, \Gamma_2)$, we know that $F_1^{*,k}$ and $G_1^{*,k}$ are uniformly Lipschitz in $(P_{1}, \Lambda_1)$ and $(P_1,\Gamma_1)$, respectively. As a consequence, the following BSDE
\begin{align*}
\begin{cases}
dP_{1,t}^k=-\Big[2rP_1^k+F_1^{*,k}(t,P_1^k,\Lambda_{1}^k)+G_1^{*,k}(P_1^k,\Gamma_1^k,P_2,\Gamma_2)\Big]dt
+(\Lambda_{1}^k)^{\top}dW+\int_{\R_+}\Gamma_1^k(y)\tilde \gamma(dt,dy),\\
P_{1,T}^k=1,
\end{cases}
\end{align*}
has Lipschitz generator. So by \cite[Lemma 2.4]{TL}, it admits a unique solution $(P_1^k,\Lambda_1^k,\Gamma_1^k)\in S^{2}_{\mathbb{F}}(0, T;\mathbb{R})\times L^{2}_{\mathbb{F}}(0, T;\mathbb{R}^n)\times \ltwonu$.

It is easily seen that 
\begin{align}\label{comparisonkey}
&\quad\;\; G_1^{*,k}(P,\Gamma,P_2,\Gamma_2)-G_1^{*,k}(P,\tilde \Gamma,P_2,\Gamma_2)\nn\\
&\geq \inf_{0\leq u\leq k} \big[ G_1(u,P,\Gamma,P_2,\Gamma_2)
- G_1(u,P,\tilde\Gamma,P_2,\Gamma_2)\big]\nn\\
&=\inf_{0\leq u\leq k} \int_{\R_+} (\Gamma(y)-\tilde\Gamma(y))\big[[(1-uy)^+]^2-1\big] \lambda \nu(dy)\nn\\
&\geq \int_{\R^+}\rho(y, \Gamma(y),\tilde\Gamma(y))(\Gamma(y)-\tilde\Gamma(y))\lambda \nu(dy),
\end{align}
where
\begin{align*}
\rho(y, \Gamma,\tilde\Gamma)&=\inf_{0\leq u\leq k}\big[[(1-uy)^+]^2-1\big]\mathbf{1}_{\Gamma\geq\tilde\Gamma}+\sup_{0\leq u\leq k}\big[[(1-uy)^+]^2-1\big]\mathbf{1}_{\Gamma<\tilde\Gamma}\\
&=\big[[(1-ky)^+]^2-1\big]\mathbf{1}_{\Gamma\geq\tilde\Gamma}, \ \forall\; y\in\R_+.
\end{align*}
It is critical to notice that $$-1\leq\rho(y, \Gamma,\tilde\Gamma)\leq 0.$$ This allows us to apply the comparison theorem for BSDEs with jumps (see, e.g., Quenes and Sulem \cite[Theorem 4.2]{QS}) to get $P^k_{1}\geq P^{k+1}_{1}$, i.e., $P^k_1$ is decreasing in $k$.

We next prove that $P_1^k$ is uniformly positive lower and upper bounded for all $k$.
Applying It\^{o}'s formula to $P_{1,t}^ke^{\int_0^t2r_sds}$ and noticing $r$ is deterministic, we get
\begin{align}\label{upperbound}
P_{1,t}^k& =e^{\int_t^T 2r_udu}+\E\Big[\int_t^T e^{\int_t^s 2r_udu }\Big(F_1^{*,k}(s,P_1^k,\Lambda_{1}^k)+G_1^{*,k}(s,P_1^k,\Gamma_1^k,P_2,\Gamma_2)\Big)ds\;\Big|\;\mathcal{F}_t\Big]\nn\\
&\leq e^{\int_t^T 2r_udu}\leq e^{\int_0^T 2|r_u|du},
\end{align}
where the first inequality is due to that
$$F_1^{*,k}(s,P_1^k,\Lambda_{1}^k)\leq F_1^{k}(s,0,P_1^k,\Lambda_{1}^k)=0,$$
and
$$G_1^{*,k}(P_1^k,\Gamma_1^k,P_2,\Gamma_2)\leq G_1^{k}(0,P_1^k,\Gamma_1^k,P_2,\Gamma_2)=0.$$
Let $c_1>0$ be a large constant such that
\begin{align}
2r-\mu^{\top}(\sigma\sigma^{\top})^{-1}\mu-\lambda\geq -c_1.
\end{align}
Because $(\underline P, \underline\Lambda, \underline\Gamma)=(e^{-c_1(T-t)},0,0)
\in S^{2}_{\mathbb{F}}(0, T;\mathbb{R})\times L^{2}_{\mathbb{F}}(0, T;\mathbb{R}^n)\times L^{2,\nu}_{\mathcal{P}}(0,T,\R)$
satisfies the following BSDE:
\begin{align*}
\underline P_t=1&-\int_t^T c_1\underline Pdt-\int_t^T\underline \Lambda^{\top}dW-\int_t^T\int_{\R_+}\underline\Gamma(y)\tilde \gamma(dt,dy),
\end{align*}
and
\begin{align*}
&\quad 2r\underline P+F_1^{*,k}(t,\underline P,\underline\Lambda)+G_1^{*,k}(\underline P,\underline\Gamma,P_2,\Gamma_2)\nn\\
&\geq 2r\underline P+\inf_{v\in\R^m}F_1(t,v,\underline P,\underline \Lambda)+\inf_{u\geq 0} G_1(u,\underline P,\underline \Gamma,P_2,\Gamma_2)\nn\\
&\geq\Big(2r-\mu^{\top}(\sigma\sigma^{\top})^{-1}\mu-\lambda\Big)\underline P\geq-c_1\underline P,
\end{align*}
we apply the comparison theorem \cite[Theorem 4.2]{QS} again to get
\begin{align}
P_{1,t}^k\geq \underline P_t=e^{-c_1(T-t)}\geq e^{-c_1T}.
\end{align}
We have now established uniform positive lower and upper bounds $\vartheta\leq P_1^k\leq M$ for all $k$ with some constants $M>\vartheta>0$.

Denote by $(T_n)_{n\in\mathbb{N}}$ the increasing sequence of jump times of the Poisson process $\{N_t\}$. We have
\begin{align*}
\E\Big[\int_0^T\int_{\R_+}\mathbf{1}_{\{P_{1,t-}^k+\Gamma_{1,t}^k(y)<\vartheta\}}\lambda \nu(dy)dt\Big]
&=\E\Big[\int_0^T\int_{\R_+}\mathbf{1}_{\{P_{1,t-}^k+\Gamma_{1,t}^k(y)<\vartheta\}}\gamma(dt,dy)\Big]\\
&=\E\Big[\sum_{n\in\mathbb{N},T_n\leq T}\mathbf{1}_{\{P_{1,T_n-}^k+\Gamma_{1,T_n}^k(-\Delta R_{T_n})<\vartheta\}}\Big]\\
&=\E\Big[\sum_{n\in\mathbb{N},T_n\leq T}\mathbf{1}_{\{P_{1,T_n}^k<\vartheta\}}\Big]=0,
\end{align*}
where $R_t$ is given by \eqref{Rprocess} and $\Delta R_{T_n}= R_{T_n}- R_{T_n-}$. Hence,
\begin{align}\label{PGammageq}
P_{1,t-}^k+\Gamma_{1,t}^k\geq\vartheta. 
\end{align} 
Similarly, we can establish 
\begin{align}\label{PGammaleq}
P_{1,t-}^k+\Gamma_{1,t}^k\leq M. 
\end{align}
Now $$\vartheta-M\leq \vartheta- P_{1,t-}^k\leq \Gamma_{1,t}^k\leq M- P_{1,t-}^k\leq M-\vartheta, 
$$
therefore, the sequence $\Gamma_1^k$, $k=1,2,\cdots,$ belongs to $\linnu $ and is uniformly bounded by $M$.

Thanks to the monotonicity, we can define $P_{1,t}:=\lim_{k\rightarrow\infty}P_{1,t}^k$. Since both $\vartheta$ and $M$ are independent of $k$,
\[
\vartheta\leq P_{1,t}\leq M, \ t\in[0,T].
\]
Applying It\^{o}'s formula to $(P_{1,t}^k)^2$, we deduce that
\begin{align*}
\begin{cases}
d(P_{1,t}^k)^2=\Big\{-2P_{1}^{k}\Big[2rP_{1}^{k}+F_1^{*,k}(t,P_1^k,\Lambda_{1}^k)+G_1^{*,k}(P_1^k,\Gamma_1^k,P_2,\Gamma_2)\Big] \\
\qquad\qquad\qquad\qquad
+|\Lambda_{1}^k|^2+\int_{\R_+}\Gamma_{1}^k(y)^2\lambda \nu(dy)\Big\}dt\\
\qquad\qquad+2P_{1}^k(\Lambda_{1}^k)^{\top}dW+\int_{\R_{+}}[(P_{1,t-}^k+\Gamma_{1,t}^k(y))^2-(P_{1,t-}^k)^2]\tilde \gamma(dt,dy)\\
(P_{1,T}^k)^2=1.
\end{cases}
\end{align*}
Taking expectation on both sides, noting $\vartheta\leq P_{1}^k\leq M$, $F_1^{*,k}\leq0$ and $G_1^{*,k}\leq 0$, we have
\begin{align}\label{L2bound}
&\quad (P_{1,0}^k)^2+\E\int_0^T|\Lambda_{1}^k|^2ds+\E\int_0^T\int_{\R_{+}}\Gamma_{1}(y)^2\lambda \nu(dy)ds\nn\\
&=1+\E\int_0^T2P_{1}^k\Big[2rP_{1}^k+F_1^{*,k}(t,P_1^k,\Lambda_{1}^k)+G_1^{*,k}(P_1^k,\Gamma_1^k,P_2,\Gamma_2)\Big]ds\nn\\
&\leq 1+\E\int_0^T4r(P_{1}^k)^2ds\nn\\
&\leq 1+4M^2 \int_0^T|r|ds.
\end{align}
Therefore, the sequence $(\Lambda_{1}^k, \Gamma_{1}^k)$, $k=1,2,\cdots,$ is bounded in $L^2_{\mathbb{F}}(0,T;\mathbb{R}^n)\times \ltwonu$, thus we can extract a subsequence (that we still respectively denote by $(\Lambda_{1}^k, \Gamma_{1}^k)$) converging in the weak sense to some $(\Lambda_{1}, \Gamma_{1})\in L^2_{\mathbb{F}}(0,T;\mathbb{R}^n)\times \ltwonu$.

Now it is standard (see, e.g., Antonelli and Mancini \cite{AM}, Kohlmann and Tang \cite{KT}) to show the strong convergence
\[\lim_{k\rightarrow\infty}\E\int_0^T|\Lambda_1^k-\Lambda_1|^2dt=0, \ \lim_{k\rightarrow\infty}\E\int_0^T\int_{\R_+}|\Gamma_1^k-\Gamma_1|^2\lambda\nu(dy)dt=0,\]
and consequently, we can conclude $(P_1,\Lambda_1,\Gamma_1)$ is a uniformly positive solution to the BSDE \eqref{P1}. A complete proof is given in Appendix \ref{appnB} for the readers' convenience.
\bigskip

We now turn to the proof of the uniqueness. Suppose $(P_1,\Lambda_1,\Gamma_1),(\tilde P_1,\tilde\Lambda_1,\tilde\Gamma_1)\in S^{\infty}_{\mathbb{F}}(0, T;\mathbb{R})\times L^{2}_{\mathbb{F}}(0, T;\mathbb{R}^n)\times \linnu $ are two uniformly positive solutions of \eqref{P1} such that $$\vartheta\leq P_1,\tilde P_1\leq M,\
\vartheta\leq P_{1,t-}+\Gamma_{1,t},\ \tilde P_{1,t-}+\tilde\Gamma_{1,t}\leq M$$ for some constants $M>\vartheta>0$. Since $(P_2,\Lambda_2,\Gamma_2)\in S^{\infty}_{\mathbb{F}}(0, T;\mathbb{R})\times L^{2}_{\mathbb{F}}(0, T;\mathbb{R}^n)\times \linnu $ is a uniformly positive solution to \eqref{P2}, we may also assume $$\vartheta\leq P_{2,t},\ P_{2,t-}+\Gamma_{2,t}\leq M.$$

Denote $\varrho:=\frac{\vartheta}{M}$, then $0<\varrho< 1$. Set
$$(U,V,\Phi)=\Big(\ln P_1,\frac{\Lambda_1}{P_1},\ln\Big(1+\frac{\Gamma_1}{P_{1,t-}}\Big)\Big)$$ and $$(\tilde U,\tilde V,\tilde \Phi)=\Big(\ln \tilde P_1,\frac{\tilde \Lambda_1}{\tilde P_1},\ln\Big(1+\frac{\tilde \Gamma_1}{\tilde P_{1,t-}}\Big)\Big).$$
Then
\begin{align}\label{Philowerbound}
\varrho\leq e^{\Phi}, e^{\tilde\Phi}\leq \varrho^{-1}, \
\frac{P_{2,t-}+\Gamma_{2,t}}{e^{U_{t-}}}, \frac{P_{2,t-}+\Gamma_{2,t}}{e^{\tilde U_{t-}}}\geq \varrho,
\end{align}
and they satisfy the following BSDE with jumps:
\begin{align*}
\begin{cases}
d U=-\Big\{2r+\tilde F^*(V)+\tilde G^*(U,\Phi)+\frac{1}{2}|V|^2+\int_{\R_+}(e^\Phi-\Phi-1)\lambda \nu(dy)\Big\}dt\\
\quad\quad~+V^{\top}dW+\int_{\R_+}\Phi\tilde\gamma(dt,dy)\\
U_T=0,
\end{cases}
\end{align*}
where
\begin{align*}
\tilde F(t,\omega,v,V)&:=|\sigma^{\top}v|^2+2v^{\top}(\mu+\sigma V),\\
\tilde G(t,u,U,\Phi)&:=\int_{\R_+}\Big(e^\Phi\big[[(1-uy)^+]^2-1\big]+\frac{P_{2,t-}
+\Gamma_{2,t}}{e^{U_{t-}}}[(1-uy)^-]^2\Big)\lambda \nu(dy)\\
&\quad\;+2u(b+\lambda b_Y),\\
\tilde F^*(t,\omega,V)&:=\inf_{v\in\Pi}\tilde F(t,\omega,v,V),\\
\tilde G^*(t,U,\Phi)&:=\inf_{u\geq0}\tilde G(t,u,U,\Phi).
\end{align*}
Set $\bar U=U-\tilde U$, $\bar V=V-\tilde V$, $\bar \Phi=\Phi-\tilde\Phi$,
applying It\^{o}'s formula to $\bar U^2$, we deduce that
\begin{align*}
&\quad\;\bar U_t^2+\int_t^T|\bar V|^2ds+\int_t^T\int_{\R_+}\bar\Phi^2\lambda \nu(dy)ds\\
&=\int_t^T2\bar U
\Big[\tilde F^*(V)-\tilde F^*(\tilde V)+\tilde G^*(U,\Phi)-\tilde G^*(\tilde U, \tilde \Phi)+\frac{1}{2}(V+\tilde V)'\bar V\\
&\qquad+\int_{\R_+}\Big((e^{\Phi}-\Phi-1)-(e^{\tilde \Phi}-\tilde \Phi-1)\Big)\lambda \nu(dy)\Big]ds\\
&\quad
-\int_t^T2\bar U\bar V^{\top}dW-\int_t^T\int_{\R_+}(2\bar U_{t-}\bar\Phi+\bar\Phi^2)\tilde \gamma(ds,dy).
\end{align*}

Notice $\tilde F^*(t,V)\leq\tilde F(t,0,V)=0$ and
\begin{align*}
\tilde F(t,v,V)\geq \delta|v|^2-c_1(1+|v||V|)>0,
\end{align*}
if $|v|>c_1(1+|V|)$ for some large constant $c_1>0$. So
\begin{align*}
\tilde F^*(t,V):=\inf_{\substack{v\in\Pi\\|v|\leq c_1(1+|V|)}}\tilde F(t,v,V),
\end{align*}
which yields
\begin{align*}
|\tilde F^*(V)-\tilde F^*(\tilde V)|\leq \sup_{\substack{v\in\Pi\\|v|\leq c_1(1+|V|+|\tilde V|)}}|2v\sigma\bar V|.
\end{align*}

A similar analysis to the proof of \eqref{L2bound} shows that
\[
\esssup_{\tau}\;\E\Big[\int_{\tau}^T(|\Lambda_{1,s}|^2+|\tilde\Lambda_{1,s}|^2)ds\Big|\mathcal{F}_{\tau}\Big]<\infty.
\]
Hence according to Lemma \ref{A4}, $\Lambda_1,\tilde\Lambda_1\in\BMO$. Recall that $P_1,\tilde P_1\geq\vartheta$ are uniformly bounded, hence $V,\ \tilde V\in\BMO$. And we can define $\beta\in\BMO$ in an obvious way such that
\begin{align*}
\tilde F^*(V)-\tilde F^*(\tilde V)+\frac{1}{2}(V+\tilde V)\bar V=\beta^{\top}\bar V,
\end{align*}
and for some constant $c_2>0$,
\[
|\beta|\leq c_2(1+|V|+|\tilde V|).
\]
According to Lemma \ref{A5}, the Dol$\acute{\mathrm{e}}$ans-Dade stochastic exponential $\mathcal{E}\big(\int_0^t\beta^{\top}dW\big)$ is a uniformly integrable martingale. From Girsanov's theorem, $$W^{\beta}:=W-\int_0^t\beta_sds$$ is a Brownian motion under the probability $\mathbb{Q}$ defined by
\begin{align*}
\frac{d\mathbb{Q}}{d\mathbb{P}}\Bigg|_{\mathcal{F}_T}=\mathcal{E}\Big(\int_0^T\beta_s^{\top}dW_s\Big).
\end{align*}
Then
\begin{align}\label{barUsquare}
&\quad\;\bar U_t^2+\int_t^T|\bar V|^2ds+\int_t^T\int_{\R_+}\bar\Phi^2\lambda \nu(dy)ds\nn\\
&=\int_t^T2\bar U\Big[\tilde G^*(U,\Phi)-\tilde G^*(\tilde U,\Phi)+\tilde G^*(\tilde U,\Phi)-\tilde G^*(\tilde U, \tilde \Phi)\nn\\
&\qquad\qquad+\int_{\R_+}\Big((e^{\Phi}-\Phi-1)-(e^{\tilde \Phi}-\tilde \Phi-1)\Big)\lambda \nu(dy)\Big]ds\nn\\
&\quad-\int_t^T2\bar U\bar V^{\top}dW^{\beta}-\int_t^T\int_{\R_+}(2\bar U_{t-}\bar\Phi+\bar\Phi^2)\tilde \gamma(dt,dy)\nn\\
&\leq\int_t^T2\bar U
\Big[\tilde G^*(\tilde U,\Phi)-\tilde G^*(\tilde U, \tilde \Phi)\nn\\
&\quad+\int_{\R_+}\Big((e^{\Phi}-\Phi-1)-(e^{\tilde \Phi}-\tilde \Phi-1)\Big)\lambda \nu(dy)\Big]ds\nn\\
&\quad
-\int_t^T2\bar U\bar V^{\top}dW^{\beta}-\int_t^T\int_{\R_+}(2\bar U_{t-}\bar\Phi+\bar\Phi^2)\tilde \gamma(dt,dy),
\end{align}
where the inequality is due to that the map $U\mapsto\tilde G^*(U,\Phi)$ is decreasing for each $\Phi\in L^{2, \nu}$.

Because $U,\Phi,\tilde U,\tilde \Phi$ are bounded and \eqref{Philowerbound}, we have
\begin{align*}
\tilde G(u,U,\Phi)
&=\int_{\R_+}\Big[e^\Phi[(1-uy)^+]^2+\frac{P_{2,t-}+\Gamma_{2,t}}{e^{U_{t-}}}[(1-uy)^-]^2 \Big]\lambda\nu(dy)\\
&\quad\;+2u(b+\lambda b_Y)-\int_{\R_+}e^{\Phi}\lambda\nu(dy)\\
&\geq \varrho\int_{\R_+}(u^2y^2 -2uy+1)\lambda \nu(dy)+2u(b+\lambda b_Y)-\lambda \varrho^{-1}\\
&=\varrho\lambda\sigma_Y^2 u^2 -2u(\varrho\lambda b_Y-(b+\lambda b_Y))+\varrho\lambda-\lambda \varrho^{-1}>0,
\end{align*}
if $u>c_3$ for some large constant $c_3>0$. In view of $$\tilde G^*(U,\Phi)\leq \tilde G(0,U,\Phi)=0,$$ we have
\begin{align*}
\tilde G^*(U,\Phi):=\inf_{0\leq u\leq c_3}\tilde G(u,U,\Phi).
\end{align*}
Notice that $\Phi,\tilde\Phi$ are $\nu$-essentially bounded, hence, for some large constant $c_4>0$,
\begin{align*}
|\tilde G^*(\tilde U,\Phi)-\tilde G^*(\tilde U, \tilde \Phi)|
&\leq \sup_{0\leq u\leq c_3}\Big|\int_{\R_+}(e^\Phi-e^{\tilde\Phi})\big[[(1-uy)^+]^2-1\big]\lambda \nu(dy)\Big|\\
&\leq 2\int_{\R_+}\Big|e^\Phi-e^{\tilde\Phi}\Big|\lambda \nu(dy)\\
&\leq c_4\int_{\R_+}|\bar\Phi|\lambda \nu(dy),
\end{align*}
and
\begin{align*}
\Big|(e^{\Phi}-\Phi-1)-(e^{\tilde \Phi}-\tilde \Phi-1)\Big|\leq c_4|\bar\Phi|.
\end{align*}
Taking expectation w.r.t. $\mathbb{Q}$ on both side of \eqref{barUsquare}, we obtain
\begin{align*}
&\quad\bar U_t^2+\E^{\mathbb{Q}}_t\Big[\int_t^T|\bar V|^2ds+\int_t^T\int_{\R_+}\bar\Phi^2\lambda \nu(dy)ds\Big]\\
&\leq c_4\E^{\mathbb{Q}}_t\Big[\int_t^T 2\bar U\int_{\R_+}|\bar\Phi|\lambda \nu(dy)ds\Big]\\
&\leq \E^{\mathbb{Q}}_t\Big[\int_t^T2 c_4^2\lambda\bar U^2ds\Big]+\frac{1}{2\lambda}\E^{\mathbb{Q}}_t\Big[\int_t^T\Big(\int_{\R_+}|\bar\Phi|\lambda \nu(dy)\Big)^2ds\Big]\\
&\leq \E^{\mathbb{Q}}_t\Big[\int_t^T2 c_4^2\lambda\bar U^2ds\Big]+\frac{1}{2\lambda}\E^{\mathbb{Q}}_t\Big[\int_t^T\Big(\int_{\R_+}|\bar\Phi|^2\lambda \nu(dy)\Big)\Big(\int_{\R_+}\lambda \nu(dy)\Big)ds\Big]\\
&= \E^{\mathbb{Q}}_t\Big[\int_t^T 2c_4^2\lambda\bar U^2ds\Big]+\frac{1}{2}\E^{\mathbb{Q}}_t\Big[\int_t^T\int_{\R_+}\bar\Phi^2\lambda \nu(dy)ds\Big].
\end{align*}
From Gronwall's inequality and the right continuity of $\bar U$, we get $\bar U=0$. Consequently, $\bar V=\bar\Phi=0$ follows from above. This completes the proof of the uniqueness.
\eof

\begin{theorem}\label{Th:P2}
The BSDE \eqref{P2} admits a unique uniformly positive solution $(P_2,\Lambda_2,\Gamma_2)\in S^{\infty}_{\mathbb{F}}(0, T;\mathbb{R})\times L^{2}_{\mathbb{F}}(0, T;\mathbb{R}^n)\times \linnu $.
\end{theorem}
\pf
We will use the same notations as in the proof of Theorem \ref{Th:P1}. It is easy to check that
\begin{align*}
G_2^*(P_2,\Gamma_2)&=\inf_{u\geq0} G_2(u,P_2,\Gamma_2),
\end{align*}
where
\begin{align}\label{G2another}
G_2(u,P_2,\Gamma_2)
&=u^2\int_{\R_+}(P_2+\Gamma_2)y^2\lambda \nu(dy)+2u\Big[\int_{\R_+}(P_2+\Gamma_2)y\lambda\nu(dy)-P_2(b+\lambda b_Y)\Big]\nn\\
&=\int_{\R_+}(P_2+\Gamma_2)\big[[(1+uy)^+]^2-1\big]\lambda \nu(dy)-2uP_2(b+\lambda b_Y).
\end{align}
It gives
\begin{align*}
G_2^{*,k}(P_2,\Gamma_2)-G_2^{*,k}(P_2,\tilde\Gamma_2)
&\geq \inf_{0\leq u\leq k} \big[G_2(u,P_2,\Gamma_2)-G_2(u,P_2,\tilde\Gamma_2)\big]\nn\\
&=\inf_{0\leq u\leq k} \int_{\R_+} (\Gamma_2(y)-\tilde\Gamma_2(y))\big[u^2y^2+2uy \big] \lambda \nu(dy)\nn\\
&\geq \int_{\R^+}\rho(y, \Gamma_2(y),\Gamma_2(y))(\Gamma_2(y)-\tilde\Gamma_2(y))\lambda \nu(dy),
\end{align*}
where
\begin{align*}
\rho(y, \Gamma,\tilde\Gamma)
&= (k^2y^2+2ky)\mathbf{1}_{\Gamma<\tilde\Gamma}, \ \forall\; y\in\R_+.
\end{align*}
Replacing $\rho$ in the proof of Theorem \ref{Th:P1} by the above one, noting that\footnote{Recall that the claims $\{Y_i\}$ are assumed to be bounded, hence $y$ will actually take values in a bounded interval $[0,M]$ for a sufficiently large constant $M>0$.
Hence, $\rho(y, \Gamma,\tilde\Gamma)$ is essentially bounded under $\nu(dy)$.}
$$0\leq \rho(y, \Gamma,\tilde\Gamma)\leq k^2y^2+2ky,$$
and applying the comparison theorem
\cite[Theorem 4.2]{QS}, we get $P^k_{2}\geq P^{k+1}_{2}$. Then the remaining proof may be handled in much the same way as Theorem \ref{Th:P1}.
\eof

\begin{remark}
If $\Pi=\R^n$, then
\begin{align*}
F_1^*(t,\omega,P,\Lambda) &=F_2^*(t,\omega,P,\Lambda)=-\frac{1}{P}(P\mu+\sigma\Lambda)'(\sigma\sigma')^{-1}(P\mu+\sigma\Lambda).
\end{align*}
\end{remark}

\begin{remark}\label{detemin}
If $m=n=1$, $\Pi=\R_+$, and $r,\mu\geq0,\sigma$ are all constants,
then
\begin{align*}
&\quad\;(P_1,\Lambda_1,\Gamma_1,P_2,\Lambda_2,\Gamma_2)=\Big(e^{2r(T-t)},\; 0,\;0,\; e^{(2r-\frac{b^2}{\lambda\sigma_Y^2}-\frac{\mu^2}{\sigma^2})(T-t)},\; 0,\;0\Big)
\end{align*}
is the unique uniformly positive solution to the system of BSDEs \eqref{P1} and \eqref{P2}. This recovers the results in \cite{BZ}.
\end{remark}

\subsection{Solution to the relaxed problem \eqref{optmun}}\label{rp}

Based on the solution to the SREs \eqref{P1} and \eqref{P2}, we are now ready to provide an optimal solution to the relaxed problem \eqref{optmun}.

For $P>0$, $\Lambda\in\R^n$, define
\begin{align}\label{v1}
\hat v_1(t,\omega,P,\Lambda) &=\argmin_{v\in\Pi}F_1(t,\omega,v,P,\Lambda),\\
\hat v_2(t,\omega,P,\Lambda) &=\argmin_{v\in\Pi}F_2(t,\omega,v,P,\Lambda),\label{v2}
\end{align}
and for $P_1,P_1+\Gamma_1,P_2,P_2+\Gamma_2>0$, define
\begin{align}\label{u1}
\hat u_1(t,P_1,\Gamma_1,P_2,\Gamma_2)&=\argmin_{u\geq0} G_1(u,P_1,\Gamma_1,P_2,\Gamma_2),\\
\hat u_2(t,P_2,\Gamma_2) &
=\frac{\Big(P_2b-\int_{\R_+}\Gamma_2(y)y\lambda\nu(dy)\Big)^+}
{\int_{\R_+}(P_2+\Gamma_2(y))y^2\lambda\nu(dy)}.\label{u2}
\end{align}

\begin{remark}\label{compawithBai}
If $\Gamma_1=\Gamma_2\equiv0$, which holds if $\mu,\sigma$ are predictable only with the filtration associate with the Brownian motion $W$, or deterministic functions of $t$, then $G_1^*(P_1,0,P_2,0)=0$ and $\hat u_1=0$ is the argument minimum, which coincides with \cite[(2.16)]{BZ}. In fact, by taking partial derivative, we have
\begin{align*}
\frac{\partial G_1(u,P_1,0,P_2,0)}{\partial u}
&=-2\int_{\R_+}P_1y(1-uy)^+\lambda \nu(dy)+2\int_{\R_+}P_2y(1-uy)^-\lambda \nu(dy)\\
&\quad\; +2P_1(b+\lambda b_Y)
\end{align*}
which is nondecreasing w.r.t $u$. Since
\begin{align*}
\frac{\partial G_1(u,P_1,0,P_2,0)}{\partial u}\Big|_{u=0}
&=2P_1(b+\lambda b_Y)-2\int_{\R_+}P_1y\lambda \nu(dy)=2bP_1\geq0,
\end{align*}
we have for all $u\geq 0$,
\begin{align*}
\frac{\partial G_1(u,P_1,0,P_2,0)}{\partial u}\geq \frac{\partial G_1(u,P_1,0,P_2,0)}{\partial u}\bigg|_{u=0}\geq 0.
\end{align*}
Hence, $G_1(u,P_1,0,P_2,0)$ is nondecreasing w.r.t. $u\geq 0$. Therefore
\begin{align*}
G_1^*(P_1,0,P_2,0) =\inf_{u\geq0} G_1(u,P_1,0,P_2,0)
=G_1(0,P_1,0,P_2,0)=0.
\end{align*}
But in our model, $\Gamma_1\neq0$, $\Gamma_2\neq0$, the argument minimum $\hat u_1$ is not zero, which is apparent different from the case with determinist coefficients studied in \cite{BZ} and \cite{BG}.
\end{remark}

\begin{theorem}\label{verifi}
Let $(P_1,\Lambda_{1},\Gamma_1,P_2,\Lambda_{2},\Gamma_2)$ be the unique uniformly positive solution to the system of BSDEs \eqref{P1} and \eqref{P2}. Set
\begin{align}\label{def:h}
h_t^\zeta=\zeta e^{-\int_t^Tr_sds}-a\int_t^Te^{-\int_t^sr_{u}du}ds.
\end{align}
Then the optimal feedback investment-reinsurance strategy for the relaxed problem \eqref{optmun} is
\begin{align}
\hat\pi(t,X) &=\hat v_1(t,P_{1,t-},\Lambda_{1,t})(X_{t-}-h_t^\zeta)^++\hat v_2(t,P_{2,t-},\Lambda_{2,t})(X_{t-}-h_t^\zeta)^-,\label{hatpi}\\
\hat q(t,X)&=\hat u_1(t,P_{1,t-},\Gamma_{1,t},P_{2,t-},\Gamma_{2,t})(X_{t-}-h_t^\zeta)^+ + \hat u_2(t,P_{2,t-},\Gamma_{2,t})(X_{t-}-h_t^\zeta)^-.\label{hatq}
\end{align}
And the optimal value of \eqref{optmun} is
\begin{align}\label{valueJ}
J(\zeta) 
&=P_{1,0}[(x-h_0^\zeta)^+]^2+P_{2,0}[(x-h_0^\zeta)^-]^2-(\zeta-z)^{2}.
\end{align}
\end{theorem}

Before proving this theorem, we first prove the following lemma.
\begin{lemma}
The feedback investment-reinsurance strategy $(\hat\pi,\hat q)$ defined in \eqref{hatpi} and \eqref{hatq} is admissible, i.e.,
$(\hat\pi,\hat q)\in\mathcal{U}$.
\end{lemma}
\pf
In the following, we will use $\hat v_1, \hat v_2, \hat u_1, \hat u_2$ and $h$ to stand for $\hat v_1(t,P_{1},\Lambda_{1})$, $\hat v_2(t,P_{2},\Lambda_{2})$, $\hat u_1(t,P_{1},\Gamma_{1},P_{2},\Gamma_{2})$, $\hat u_2(t,P_{2},\Gamma_{2})$ and $h^\zeta$ respectively for notational simplicity.
By definition, it is clear that $\hat v_1, \hat v_2\in\Pi$, and $\hat u_1, \hat u_2\geq 0$, so
$\hat\pi\in\Pi$, $\hat q\geq 0$. It is only left to show that $(\hat\pi,\hat q)\in L^2_{\mathbb{F}}(0,T;\R^{m+1})$.

Substituting $\hat\pi$ and $\hat q$ into the wealth process \eqref{wealth},
\begin{align}\label{hatX}
&\quad\;d(X_t-h_t)\nn\\
&=(X_{t-}-h_t)^+\Big[(r+\hat v_1^{\top}\mu+\hat u_1b+\hat u_1\lambda b_Y)dt+\hat v_1^{\top}\sigma dW-\hat u_1\int_{\R_+}y \gamma(dt,dy)\Big]\nn\\
&\quad+ (X_{t-}-h_t)^-\Big[(-r+\hat v_2^{\top}\mu+\hat u_2 b+\hat u_2\lambda b_Y)dt+\hat v_2^{\top}\sigma dW-\hat u_2\int_{\R_+}y \gamma(dt,dy)\Big].
\end{align}
By the proof of Theorem \ref{Th:P1}, we know that $$|\hat v_1|\leq c(1+|\Lambda_1|),\ |\hat v_2|\leq c(1+|\Lambda_2|),$$ and $\hat u_1, \hat u_2$ are bounded. By \cite[basic theorem]{Ga}, the SDE \eqref{hatX} has a unique strong solution. Actually, the solution of \eqref{hatX} is given by
\begin{align}\label{explicit}
&\quad\;X_t-h_t\nn\\
&=(x-h_0)^+L_{0,t}\prod_{0<s\leq t}\Big(1-\int_{\R_+}\hat u_1 y\gamma(\{s\},dy)\Big)\mathbf{1}_{t<\tau}\nn\\
&\quad\;+(x-h_0)^+L_{0,\tau}\prod_{0<s\leq \tau}\Big(1-\int_{\R_+}\hat u_1 y\gamma(\{s\},dy)\Big)M_{\tau,t}\prod_{\tau<s\leq t}\Big(1+\int_{\R_+}\hat u_2 y\gamma(\{s\},dy)\mathbf{1}_{t\geq\tau}\Big)\nn\\
&\quad\;-(x-h_0)^-M_{0,t}\prod_{0<s\leq t}\Big(1+\int_{\R_+}\hat u_2 y\gamma(\{s\},dy)\Big),
\end{align}
where
\begin{align*}
L_{s,t}&:=e^{\int_s^t (r+\hat v_1^{\top}\mu+\hat u_1(b+\lambda b_Y)-\frac{1}{2}|\sigma^{\top}\hat v_1|^2)d\alpha+\int_s^t\hat v_1^{\top}\sigma dW},\\
M_{s,t}&:=e^{\int_s^t (r-\hat v_2^{\top}\mu-\hat u_2 (b+\lambda b_Y)-\frac{1}{2}|\sigma^{\top}\hat v_2|^2)d\alpha-\int_s^t\hat v_2^{\top}\sigma dW},\\
\tau&:=\inf\Big\{t\geq0:1-\int_{\R_+}\hat u_1y\gamma(\{t\},dy)\leq0\Big\}.
\end{align*}

Applying It\^{o}-Tanaka's formula to $P_{1,t}[(X_t-h_t)^+]^2+P_{2,t}[(X_t-h_t)^-]^2$, we obtain\footnote{Please refer to \eqref{Xhposi} and \eqref{Xhnega} for more details.}
\begin{align}\label{P1P22}
&\quad\;P_{1,t}[(X_t-h_t)^+]^2+P_{2,t}[(X_t-h_t)^-]^2\nn\\
&=P_{1,0}[(x-h_0)^+]^2+P_{2,0}[(x-h_0)^-]^2
\nn\\
&\quad+\int_0^t[(X_{s-}-h_s)^+]^2\big[\Lambda_{1,s}+2P_{1,s}\sigma^{\top}\hat v_1\big]^{\top}dW\nn\\
&\quad+\int_0^t[(X_{s-}-h_s)^-]^2\big[\Lambda_{2,s}-2P_{2,s}\sigma^{\top}\hat v_2\big]^{\top}dW\nn\\
&\quad+\int_0^t\int_{\R_+}\Big\{(P_{1,s-}+\Gamma_{1,s})\Big[[(X_{s-}-h_s-qy)^+]^2-[(X_{s-}-h_s)^+]^2\Big]\nn\\
&\quad\qquad\qquad+[(X_{s-}-h_s)^+]^2\Gamma_{1,s}\Big\}\tilde \gamma(ds,dy)\nn\\
&\quad+\int_0^t\int_{\R_+}\Big\{(P_{2,s-}+\Gamma_{2,s})\Big[[(X_{s-}-h_s-qy)^-]^2-[(X_{s-}-h_s)^-]^2\Big]\nn\\
&\quad\qquad\qquad+[(X_{s-}-h_s)^-]^2\Gamma_{2,s}\Big\}\tilde \gamma(ds,dy).
\end{align}
For any positive integer $j$, define the stoping time $\tau_j$ as follows:
\begin{align*}
\tau_j:=\inf\Big\{t\geq0\colon\int_0^t\big(|X_{s-}|^2&+|\Lambda_{1,s}|^2+|\Lambda_{2,s}|^2
+|\hat v_1(s)|^2+|\hat v_2(s)|^2\big)ds> j\Big\}\wedge T.
\end{align*}
It is obvious that $\tau_j\uparrow T$ a.s. as $j\uparrow\infty$. Notice $h$ is bounded, so by taking expectation on both sides of \eqref{P1P22}, we get
\begin{align*} 
\E\Big[P_{1,\iota\wedge\tau_j}[(X_{\iota\wedge\tau_j}-h_{\iota\wedge\tau_j})^+]^2+P_{2,\iota\wedge\tau_j}[(X_{\iota\wedge\tau_j}-h_{\iota\wedge\tau_j})^-]^2\Big]
=P_{1,0}[(x-h_0)^+]^2+P_{2,0}[(x-h_0)^-]^2,
\end{align*}
for any stopping time $\iota\leq T$.
By Theorems \ref{Th:P1} and \ref{Th:P2}, $P_1,P_2\geq\vartheta>0$. Then it follows from above that
\begin{align*}
\vartheta\E\Big[(X_{\iota\wedge\tau_j}-h_{\iota\wedge\tau_j})^2
\Big]
\leq &P_{1,0}[(x-h_0)^+]^2+P_{2,0}[(x-h_0)^-]^2.
\end{align*}
Sending $j\rightarrow\infty$, it follows from Fatou's lemma that
\begin{align*}
\E\Big[(X_{\iota}-h_{\iota})^2
\Big]\leq c,
\end{align*}
for any stopping time $\iota\leq T$. Since $h$ is bounded, this further implies $ X\in L^2_{\mathbb{F}}(0,T;\R)$.

Let
\begin{align*}
\theta_j:=\inf\Big\{t\geq0\colon \int_0^t\big(|\hat\pi_s|^2+ \hat q^2_s\big)ds> j\Big\}\wedge T.
\end{align*}
Applying It\^{o}'s formula to $(X_t-h_t)^2$ and using the elementary inequality $2|ab|\leq \frac{2}{\delta}|a|^2+\frac{\delta}{2}|b|^2$, we have, for some large constant $c>0$,
\begin{align*}
&\quad\;(x-h_0)^2+\E\int_0^{T\wedge \theta_j}(|\hat\pi^{\top}\sigma|^2+\lambda\sigma_Y^2 \hat q^2)dt\\
&=\E[(X_{T\wedge \theta_j}-h_{T\wedge \theta_j})^2]-\E\int_0^{T\wedge \theta_j} 2(X-h)[r(X-h)+\hat\pi^{\top}\mu+b\hat q]dt\\
&\leq c+c\E\int_0^T |X-h|^2dt+\frac{\delta}{2}\E\int_0^{T\wedge \theta_j}|\hat\pi|^2dt+\frac{\lambda\sigma_Y^2}{2}\E\int_0^{T\wedge \theta_j}|\hat q|^2dt.
\end{align*}
Since $\sigma\sigma'\geq \delta I_m$, after rearrangement, it follows
\begin{align*}
\frac{\delta}{2}\E\int_0^{T\wedge \theta_j}|\hat\pi|^2dt
+\frac{\lambda\sigma_Y^2}{2}\E\int_0^{T\wedge \theta_j}|\hat q|^2dt\leq c +c\E\int_0^T |X-h|^2dt.
\end{align*}
Sending $j\to\infty$, Fatou's lemma yields
\[
(\hat\pi,\hat q)\in L^2_{\mathbb{F}}(0,T;\R^{m+1}).
\]
This completes the proof.
\eof
\bigskip

\noindent\textbf{Proof of Theorem \ref{verifi}.}\\
For any $(\pi,q)\in\mathcal{U}$, applying It\^{o}-Tanaka's formula to $P_{1,t}[(X_t-h_t)^+]^2$ and $P_{2,t}[(X_t-h_t)^-]^2$ respectively, we obtain
\begin{align}\label{Xhposi}
dP_{1,t}[(X_t-h_t)^+]^2
&=\Big\{P_{1,t-}\mathbf{1}_{\{X_{t-}-h_t> 0\}}|\sigma^{\top}\pi|^2\nn\\
&\quad+\int_{\R_+}(P_{1,t-}+\Gamma_{1,t})\Big[[(X_{t-}-h_t-qy)^+]^2-[(X_{t-}-h_t)^+]^2\Big]\lambda \nu(dy)\nn\\
&\quad+2(X_{t-}-h_t)^+[\pi^{\top}(P_{1,t-}\mu+\sigma\Lambda_{1,t})+q P_{1,t-}(b+\lambda b_Y)]\nn\\
&\quad-[F_1^*(P_{1,t-},\Lambda_{1,t})+G_1^*(P_{1,t-},\Gamma_{1,t},P_{2,t-},\Gamma_{2,t})][(X_{t-}-h_t)^+]^2\Big\}dt\nn\\
&\quad+\Big[[(X_{t-}-h_t)^+]^2\Lambda_{1,t}+2(X_{t-}-h_t)^+P_{1,t-}\sigma^{\top}\pi\Big]^{\top}dW\nn\\
&\quad+\int_{\R_+}\Big\{(P_{1,t-}+\Gamma_{1,t})\Big[[(X_{t-}-h_t-qy)^+]^2-[(X_{t-}-h_t)^+]^2\Big] \nn\\
&\quad\qquad+[(X_{t-}-h_t)^+]^2\Gamma_{1,t}\Big\}\tilde \gamma(dt,dy),
\end{align}
and
\begin{align}\label{Xhnega}
dP_{2,t}[(X_t-h_t)^-]^2
&=\Big\{P_{2,t-}\mathbf{1}_{\{X_{t-}-h_t\leq 0\}}|\sigma^{\top}\pi|^2 \nn\\ &\quad+\int_{\R_+}(P_{2,t-}+\Gamma_{2,t})\Big[[(X_{t-}-h_t-qy)^-]^2-[(X_{t-}-h_t)^-]^2\Big]\lambda \nu(dy)\nn\\
&\quad-2(X_{t-}-h_t)^-[\pi^{\top}(P_{2,t-}\mu+\sigma\Lambda_{2,t})+q P_{2,t-}(b+\lambda b_Y)]\nn\\
&\quad-[F_2^*(P_{2,t-},\Lambda_{2,t})+G_2^*(P_{2,t-},\Gamma_{2,t})][(X_{t-}-h_t)^-]^2\Big\}dt\nn\\
&\quad+\Big[[(X_{t-}-h_t)^-]^2\Lambda_{2,t}-2(X_{t-}-h_t)^-P_{2,t-}\sigma^{\top}\pi\Big]^{\top}dW\nn\\
&\quad+\int_{\R_+}\Big\{(P_{2,t-}+\Gamma_{2,t}) \Big[[(X_{t-}-h_t-qy)^-]^2-[(X_{t-}-h_t)^-]^2\Big]\nn\\
&\quad\qquad+[(X_{t-}-h_t)^-]^2\Gamma_{2,t}\Big\}\tilde \gamma(dt,dy).
\end{align}

For any positive integer $j$, define the stoping time $\tau_j$ as follows:
\begin{align*}
\tau'_j:=\inf\Big\{t\geq0\colon \int_0^t\big(|X_{s-}^{\pi,q}|^2&+|\Lambda_{1,s}|^2+|\Lambda_{2,s}|^2
+|\pi_s|^2+|q_s|^2\big)ds> j\Big\}\wedge T.
\end{align*}
Then combining \eqref{Xhposi} and \eqref{Xhnega}, integrating from $0$ to $T\wedge\tau_j$, taking expectation, we have
\begin{align}\label{geq}
&\quad\E\Big[P_{1,T\wedge\tau'_j}[(X_{T\wedge\tau'_j}-h_{T\wedge\tau'_j})^+]^2+P_{2,T\wedge\tau'_j}
[(X_{T\wedge\tau'_j}-h_{T\wedge\tau'_j})^-]^2\Big]\nn\\
&=P_{1,0}[(x-h_0)^+]^2+P_{2,0}[(x-h_0)^-]^2+\E\int_0^{T\wedge\tau'_j}\phi(t,X_{t-},\pi,q)dt,
\end{align}
where
\begin{align*}
\phi(t,X_{t-},\pi,q)
&=P_{1,t-}\mathbf{1}_{\{X_{t-}-h_t> 0\}}|\sigma^{\top}\pi|^2\nn\\
&\qquad\quad+\int_{\R_+}(P_{1,t-}+\Gamma_{1,t})\Big[[(X_{t-}-h_t-qy)^+]^2-[(X_{t-}-h_t)^+]^2\Big]\lambda \nu(dy)\nn\\
&\quad+P_{2,t-}\mathbf{1}_{\{X_{t-}-h_t\leq 0\}}|\sigma^{\top}\pi|^2\nn\\
&\qquad\quad+\int_{\R_+}(P_{2,t-}+\Gamma_{2,t})\Big[[(X_{t-}-h_t-qy)^-]^2-[(X_{t-}-h_t)^-]^2\Big]\lambda \nu(dy)\nn\\
&\quad+2(X_{t-}-h_t)^+[\pi^{\top}(P_{1,t-}\mu+\sigma\Lambda_{1,t})+q P_{1,t-}(b+\lambda b_Y)]\nn\\
&\quad-2(X_{t-}-h_t)^-[\pi^{\top}(P_{2,t-}\mu+\sigma\Lambda_{2,t})+q P_{2,t-}(b+\lambda b_Y)]\nn\\
&\quad-[F_1^*(P_{1,t-},\Lambda_{1,t})+G_1^*(P_{1,t-},\Gamma_{1,t},P_{2,t-},\Gamma_{2,t})][(X_{t-}-h_t)^+]^2\Big]\nn\\
&\quad-[F_2^*(P_{2,t-},\Lambda_{2,t})+G_2^*(P_{2,t-},\Gamma_{2,t})][(X_{t-}-h_t)^-]^2\Big].
\end{align*}
Define a pair process
\begin{align*}
(v_t,u_t)=
\begin{cases}
\Big(\frac{\pi_t}{|X_{t-}-h_t|}, \frac{q_t}{|X_{t-}-h_t|}\Big), \ & \mbox{if} \ |X_{t-}-h_t|>0;\\
(0,0), \ & \mbox{if} \ |X_{t-}-h_t|=0.
\end{cases}
\end{align*}
Since $\Pi$ is a cone, $v_t\in\Pi$ and $u_t\geq 0$. If $X_{t-}-h_t>0$, then
\begin{align*}
\phi(t,X_{t-},\pi,q)&=(X_{t-}-h_t)^2\Big\{P_{1,t-}|\sigma^{\top}v|^2 +2v^{\top}(P_{1,t-}\mu+\sigma\Lambda_{1,t})
-F_1^*(P_{1,t-},\Lambda_{1,t})\\
&\quad+\int_{\R_+}(P_{1,t-}+\Gamma_{1,t})\Big[[(1-uy)^+]^2-1\Big]\lambda \nu(dy)\\
&\quad+\int_{\R_+}(P_{2,t-}+\Gamma_{2,t})[(1-uy)^-]^2\lambda \nu(dy)\\
&\quad+2u P_{1,t-}(b+\lambda b_Y)-G_1^*(P_{1,t-},\Gamma_{1,t},P_{2,t-},\Gamma_{2,t})\Big\}\\
&\geq 0,
\end{align*}
by the definitions of $F_1^*(P_{1},\Lambda_1)$ and $G_1^*(P_{1},\Gamma_1,P_{2},\Gamma_2)$. Similarly we have $\phi(t,X_{t-},\pi,q)\geq 0$ if $X_{t-}-h_t\leq0$. %
Hence it follows from \eqref{geq} that
\begin{align*}
&\quad\;\E\Big[P_{1,T\wedge\tau'_j}[(X_{T\wedge\tau'_j-}-h_{T\wedge\tau'_j})^+]^2+
P_{2,T\wedge\tau'_j}[(X_{T\wedge\tau'_j-}-h_{T\wedge\tau'_j})^-]^2\Big]\\
&\geq P_{1,0}[(x-h_0)^+]^2+P_{2,0}[(x-h_0)^-]^2.
\end{align*}
By the standard theory of SDE, we have $$\E\Big[\sup_{t\in[0,T]}X_t^2\Big]<\infty$$ for any $(\pi,q)\in\mathcal{U}$. Letting $j\rightarrow\infty$, the dominated convergence theorem yields
\begin{align*}
\E\Big[(X_{T}-h_{T})^2\Big]\geq P_{1,0}[(x-h_0)^+]^2+P_{2,0}[(x-h_0)^-]^2,
\end{align*}
which becomes an equality under the controls \eqref{hatpi} and \eqref{hatq}. This completes the proof of Theorem \ref{verifi}.

\section{Solution to the mean-variance problem \eqref{optm}}\label{solution}
We are now ready to solve the original problem \eqref{optm}. The following lemma will be used in determining the Lagrange multiplier in \eqref{duality}.
\begin{lemma}\label{lemma}
Let $(P_1,\Lambda_{1},\Gamma_1,P_2,\Lambda_{2},\Gamma_2)$ be the unique uniformly positive solution to the system of BSDEs \eqref{P1} and \eqref{P2}. Then it holds that
\begin{align*}
P_{1,0}e^{-2\int_0^Tr_sds}-1\leq 0, \quad P_{2,0}e^{-2\int_0^Tr_tds}-1< 0.
\end{align*}
\end{lemma}
\pf
The first inequality is given by \eqref{upperbound}. Similarly, one can prove
\begin{align*}
P_{2,0}e^{-2\int_0^Tr_tds}-1\leq 0.
\end{align*}
It remains to prove the strict inequality $$P_{2,0}e^{-2\int_0^Tr_tds}-1< 0.$$ Suppose, on the contrary, $$P_{2,0}e^{-2\int_0^Tr_tds}-1= 0.$$ It then would follow from an analog of \eqref{upperbound} that $$F_2^*(t,P_2,\Lambda_2)+G_2^*(P_2,\Gamma_2)=0.$$
Since $F_2^*$, $G_2^*\leq 0$, we get $$F_2^*(t,P_2,\Lambda_2)=G_2^*(P_2,\Gamma_2)=0.$$
Therefore, the BSDE \eqref{P2} becomes
\begin{align*}
\begin{cases}
dP_2=-2rP_2dt+\Lambda_{2}^{\top}dW+\int_{\R_+}\Gamma_2(y)\tilde \gamma(dt,dy), \\
P_{2,T}=1, \\
P_{2,t}>0, \ \ P_{2,t-}+\Gamma_{2,t}>0, \ \mbox{ for all $t\in[0, T]$.}
\end{cases}
\end{align*}
Clearly, this BSDE also admits a solution $(e^{2\int_t^Tr_sds},0,0)$. By the uniqueness of its solution, we obtain $$(P_{2,t},\Lambda_{2,t},\Gamma_{2,t})=(e^{2\int_t^Tr_sds},0,0).$$ Consequently, $$G_2^*(P_2,0)=G_2^*(P_2,\Lambda_2)=0.$$ On the other hand, it follows from the definition that
\begin{align*}
G_2^*(P_2,0)&=-\frac{[(P_2 b )^+]^2}{P_2 \lambda\sigma_Y^2}=-\frac{P_2 b^2}{\lambda\sigma_Y^2}<0,
\end{align*}
leading to a contraction.
\eof

\begin{theorem}
Let $(P_1,\Lambda_{1},\Gamma_1,P_2,\Lambda_{2},\Gamma_2)$ be the unique uniformly positive solution to the system of BSDEs \eqref{P1} and \eqref{P2}, and let $\hat v_2,\hat u_2$ be defined by
\eqref{v2} and \eqref{u2}. Then
an efficient feedback strategy for the problem \eqref{optm} is
\begin{align}\label{hatpimv}
\begin{cases}
\hat\pi(t,X)=-\hat v_2(t,P_{2,t-},\Lambda_{2,t})(X_{t-}-h_t^{\hat\zeta}),\\
\hat q(t,X)=- \hat u_2(t,P_{2,t-},\Gamma_{2,t})(X_{t-}-h_t^{\hat\zeta}),
\end{cases}
\end{align}
where $\hat \zeta$ is a constant given by
\begin{align}\label{hatzeta}
\hat \zeta&:=\frac{P_{2,0}e^{-\int_0^Tr_sds}\Big(x+a\int_0^Te^{-\int_0^tr_sds}dt\Big)-z}{P_{2,0}e^{-2\int_0^Tr_sds}-1} \geq xe^{\int_0^Tr_sds}+a \int_0^Te^{\int_t^Tr_sds}dt.
\end{align}
Moreover, the efficient frontier is a half-line, determined by
\begin{align}\label{efficient}
\mathrm{Var}(X_T)&=\frac{P_{2,0}e^{-2\int_0^Tr_sds}}{1-P_{2,0}e^{-2\int_0^Tr_sds}}\Big[\E(X_T)-\Big(xe^{\int_0^Tr_sds}+a \int_0^Te^{\int_t^Tr_sds}dt\Big)\Big]^2,
\end{align}
where
$$\E(X_T)\geq xe^{\int_0^Tr_sds}+a \int_0^Te^{\int_t^Tr_sds}dt.$$
\end{theorem}
\pf
By Lemma \ref{lemma}, $\hat\zeta$ in \eqref{hatzeta} is well-defined. From \eqref{valueJ}, we have the following expression of the value function \eqref{def:J}:
\begin{align*}
J(\zeta)=
\begin{cases}
J_1(\zeta), \ \mbox{if} \ \zeta< xe^{\int_0^Tr_sds}+a \int_0^Te^{\int_t^Tr_sds}dt,\\
J_2(\zeta), \ \mbox{if} \ \zeta\geq xe^{\int_0^Tr_sds}+a \int_0^Te^{\int_t^Tr_sds}dt,
\end{cases}
\end{align*}
where
\begin{align*}
J_1(\zeta)&:=\Big[P_{1,0}e^{-2\int_0^Tr_sds}-1\Big]\zeta^2-2\Big[P_{1,0}e^{-\int_0^Tr_sds}\Big(x+a\int_0^Te^{-\int_0^tr_sds}dt\Big)-z\Big]\zeta\\
&\qquad+P_{1,0}\Big(x+a\int_0^Te^{-\int_0^tr_sds}dt\Big)^2-z^2,
\end{align*}
and
\begin{align*}
J_2(\zeta)&:=\Big[P_{2,0}e^{-2\int_0^Tr_sds}-1\Big]\zeta^2-2\Big[P_{2,0}e^{-\int_0^Tr_sds}\Big(x+a\int_0^Te^{-\int_0^tr_sds}dt\Big)-z\Big]\zeta\\
&\qquad+P_{2,0}\Big(x+a\int_0^Te^{-\int_0^tr_sds}dt\Big)^2-z^2.
\end{align*}
Since $$z\geq xe^{\int_0^Tr_sds}+a \int_0^Te^{\int_t^Tr_sds}dt,$$ we have the inequality in \eqref{hatzeta}. A tedious calculation yields
\begin{align}\label{efficient2}
J(\hat\zeta)=\max_{\zeta\in\R} J(\zeta)
=\frac{P_{2,0}e^{-2\int_0^Tr_sds}}{1-P_{2,0}e^{-2\int_0^Tr_sds}}
\Big[z-e^{\int_0^Tr_sds}\Big(x+a\int_0^Te^{-\int_0^tr_sds}dt\Big)\Big]^2.
\end{align}
This proves \eqref{efficient}.

Finally, noting $z\geq xe^{\int_0^Tr_sds}+a\int_0^Te^{\int_t^Tr_sds}dt,$ we have
\begin{align}\label{xh0}
x-h^{\hat\zeta}_0&=x+a\int_0^Te^{-\int_0^tr_sds}dt-e^{-\int_0^Tr_sds}
\frac{P_{2,0}e^{-\int_0^Tr_sds}\Big(x+a\int_0^Te^{-\int_0^tr_sds}dt\Big)-z}{P_{2,0}e^{-2\int_0^Tr_sds}-1}\leq 0.
\end{align}
Hence we obtain
\begin{align}\label{xh}
X_t-h^{\hat\zeta}_t\leq 0
\end{align}
from \eqref{hatX} or \eqref{explicit}. Therefore, \eqref{hatpimv} follows from \eqref{hatpi}.
\eof

\begin{remark}
Assume that there is only one stock and the Brownian motion $W$ is one-dimensional, i.e., $m=n=1$. In this case, the closed convex cone $\Pi$ in $\R$ must be one of the following sets: $\R_+$, $\R_-$, $\R$.
From the definitions of $F_2$, $\hat v_2$ and $\hat u_2$ in
\eqref{v2} and \eqref{u2} respectively, we have
\begin{align}\label{v20}
\hat v_2(P_{2},\Lambda_{2})=
\begin{cases}
\frac{1}{\sigma^2}\big(\mu+\frac{\sigma\Lambda_{2}}{P_2}\big)^+, \ \mbox{if} \ \ \Pi=\R_+;\\
\frac{1}{\sigma^2}\big(\mu+\frac{\sigma\Lambda_{2}}{P_2}\big)^-, \ \mbox{if} \ \ \Pi=\R_-;\\
\frac{1}{\sigma^2}\big(\mu+\frac{\sigma\Lambda_{2}}{P_2}\big), \ \ \mbox{if} \ \ \Pi=\R,
\end{cases}
\end{align}
and
\begin{align}\label{u20}
\hat u_2(P_2,\Gamma_2)
=\frac{\Big(P_2b-\int_{\R_+}\Gamma_2(y)y\lambda\nu(dy)\Big)^+}
{\int_{\R_+}(P_2+\Gamma_2(y))y^2\lambda\nu(dy)}.
\end{align}

Assume that, furthermore, $\mu,\sigma$ are deterministic functions of $t$, then $\Lambda_2\equiv 0$ and $\Gamma_2\equiv0$ by Remark \ref{compawithBai}. In this case, we have
\begin{align}\label{v200}
\hat v_2(P_{2},\Lambda_{2})=
\begin{cases}
\frac{\mu^+}{\sigma^2}, \ & \mbox{if} \ \ \Pi=\R_+;\\
\frac{\mu^-}{\sigma^2}, \ & \mbox{if} \ \ \Pi=\R_-;\\
\frac{\mu}{\sigma^2}, \ & \mbox{if} \ \ \Pi=\R.
\end{cases}
\end{align}
and noting the definitions of $b$ and $\sigma_Y$ in \eqref{def:ba} and \eqref{def:bY},
\begin{align*}
\hat u_2(P_2,\Gamma_2) &
=\frac{P_2b^+}
{\int_{\R_+}P_2y^2\lambda\nu(dy)}=\frac{b}{\lambda\sigma_{Y}^2}>0.
\end{align*}

In the case of $\Pi=\R$, we deduce from \eqref{hatpimv} and \eqref{v200} that
\begin{align}\label{pileq0}
\hat\pi(t,X)<0 \ \mbox{if and only if} \ \mu_t<0,
\end{align}
i.e., the insurer will short-sale the stock at time $t$ if the mean excess return rate $\mu_{t}$ is negative at that moment.

Using \eqref{explicit}, \eqref{hatpimv}, \eqref{xh0}, \eqref{xh} and $\hat u_2>0$, we have
\begin{align*}
\hat q(t,X)&=- \frac{b}{\lambda\sigma_Y^2}(X_{t-}-h_t^{\hat\zeta})\\
&\geq \frac{b}{\lambda\sigma_Y^2}(h_0^{\hat\zeta}-x)\exp\Big(\int_0^t (r-\hat v_2^{\top}\mu-\hat u_2 (b+\lambda b_Y)-\frac{1}{2}|\sigma^{\top}\hat v_2|^2)ds-\int_0^t\hat v_2^{\top}\sigma dW\Big).
\end{align*}
Therefore, $\hat q(t,X)>1$ if
\begin{align*}
-\int_0^t\hat v_2^{\top}\sigma dW>\ln\Big(\frac{\lambda\sigma_Y^2}{b(h_0^{\hat\zeta}-x)}\Big)-\int_0^t (r-\hat v_2^{\top}\mu-\hat u_2 (b+\lambda b_Y)-\frac{1}{2}|\sigma^{\top}\hat v_2|^2)ds,
\end{align*}
from which we conclude $\hat q(t,X)>1$ happens with strictly positive probability, since $-\int_0^t\hat v_2^{\top}\sigma dW$ is normally distributed.

In the special case $r=0$ and $a=0$, we have
\begin{align*}
P_{2,t}=\exp\Big(-\int_t^T\big(\frac{\mu_s^2}{\sigma_s^2}+\frac{b^2}{\lambda \sigma_Y^2}\big)ds\Big),
\end{align*}
so that the efficient frontier \eqref{efficient} is reduced to
\begin{align*}
\sqrt{\mathrm{Var}(X_T)}&=\frac{1}{\sqrt{\exp\Big(\int_0^T\big(\frac{\mu_s^2}{\sigma_s^2}+\frac{b^2}{\lambda \sigma_Y^2}\big)ds\Big)-1}}\big[\E(X_T)-x\big], \ \E(X_T)\geq x,
\end{align*}
which is a half-line whose slope will increase if $\mu^2$ or $b^2$ becomes smaller or $\sigma^2$ or $\sigma_Y^2$ becomes larger.

But if one of $\mu$ or $\sigma$ is a stochastic process, then $\Lambda_2\neq0$ or $\Gamma_2\neq0$. Hence $(\Lambda_2,\Gamma_2)$, compared with the determinist coefficients model \cite{BZ} and \cite{BG}, serves to reflect the randomness of the coefficients $(\mu,\sigma)$.
\end{remark}

\section{Concluding remarks}\label{cr}
In this paper, we study an optimal investment-reinsurance problem under the mean-variance criterion. The jumps in the wealth dynamic \eqref{wealth} are of special downward type so that the underling BSDEs \eqref{P1} and \eqref{P2} are partially coupled. We prove the existence and uniqueness of uniformly positive solutions to \eqref{P1} and \eqref{P2} by pure BSDE techniques, which is interesting in its own from the point of view of BSDE theory. These results are then applied to construct a feedback control, which is proved to be optimal by a verification theorem, for the relaxed problem \eqref{optmun}. Finally, we succeed in presenting the efficient strategy and efficient frontier of the original mean-variance problem.

Further research along this line can be interesting; for instance, (1) How to solve the problem \eqref{optm} when the interest rate $r$ is a stochastic process? (2) How to solve the cone-constrained mean-variance problem or more general stochastic linear quadratic control problem with general jumps? In this case, the underling system of BSDE will be fully coupled whose solvability will be challenging. (3) If the mean excess return rate $\mu$ and volatility $\sigma$ are unbounded, how to solve the mean-variance investment-reinsurance problem \eqref{optm}? Does the system of the SREs \eqref{P1} and \eqref{P2} admit a unique solution? We cannot find any results in the literature to address SREs with jumps and unbounded coefficients. We hope to solve these problems in our future research.

\appendixpage
\addappheadtotoc
\appendix

\section{Facts about continuous BMO martingales}\label{appnA}

For the readers' convenience, let us recall the definitions and some properties of continuous BMO martingales that will be used in the proof of Theorem \ref{Th:P1}; see, e.g., \cite[P25, Equation (2.1), Corollary 2.1, Definition 2.1, Theorem 2.3]{Ka}.

\begin{definition}
Let $M$ be a uniformly integrable martingale on $[0,T]$ with $M_0=0$. For $p\geq 1$, we say $M$ is in the normed linear space $\mathrm{BMO}_p$ martingale, if
\begin{align}\label{A1}
\lVert M\rVert_{\mathrm{BMO}_p}:=\sup_{\tau}\lVert \E[|M_T-M_{\tau}|^p|\mathcal{F}_{\tau}]^{1/p}\rVert_{\infty}<\infty
\end{align}
where the supremum is taken over all stoping times $\tau\leq T$.
\end{definition}

\begin{lemma}
Let $p>1$. Then there is a positive constant $C_p$ such that for any uniformly integrable martingale $M$, it holds that
\[
\lVert M\rVert_{\mathrm{BMO}_1}\leq \lVert M\rVert_{\mathrm{BMO}_p} \leq C_p \lVert M\rVert_{\mathrm{BMO}_1}.
\]
Therefore, a martingale $M\in \mathrm{BMO}_p$ if and only if $M\in \mathrm{BMO}_q$ for every $q\geq 1$.
\end{lemma}

\begin{definition}
Each martingale $M$ in $\mathrm{BMO}_1$ is called a BMO martingale.
\end{definition}
Letting $p=2$ in \eqref{A1}, we immediately have
\begin{lemma}\label{A4}
A local martingale $M_t=\int_0^t\phi_s^\top dW_s$ is a BMO martingale on $[0,T]$ if and only if
\[
\esssup_{\tau}\;\E\Big[\int_{\tau}^T|\phi_s|^2ds\Big|\mathcal{F}_{\tau}\Big]<\infty.
\]
\end{lemma}
We will use the flowing property of BMO martingales in the proof of Theorem \ref{Th:P1}.
\begin{lemma}
\label{A5}
If $\int_0^{t}\phi_s^\top dW_s$ is a BMO martingale on $[0,T]$, then the Dol\'eans-Dade stochastic exponential $\mathcal{E}\big(\int_0^{t}\phi_s^{\top}dW_s\big)$ is a uniformly integrable martingale on $[0,T]$.
\end{lemma}

\section{Supplement of the proof of Theorem \ref{Th:P1}}\label{appnB}

For positive integers $k< l$, set
\[
P_{2,1}^{k,l}:=P_{1,t}^{k}-P_{1,t}^{l}\geq0, \ \Lambda_{1,t}^{k,l}:=\Lambda_{1,t}^{k}-\Lambda_{1,t}^{l}, \ \Gamma_{1,t}^{k,l}:=\Gamma_{1,t}^{k}-\Gamma_{1,t}^{l}.
\]
Let $\kappa>0$ be a constant to be specified later, and write $$\Psi(x)=\frac{1}{\kappa}\big(e^{\kappa x}-\kappa x-1\big).$$
Applying It\^{o}'s formula to $\Psi(P_{1,t}^{k,l})$, we get
\begin{align*}
&\quad\E[\Psi(P_{1,0}^{k,l})]+\frac{1}{2}\E\int_0^T\Psi''(P_{1,t}^{k,l})|\Lambda_1^{k,l}|^2dt\\
&\qquad\qquad+\E\int_0^T\int_{\R_+}\Big[\Psi(P_{1,t-}^{k,l}+\Gamma_{1,t}^{k,l})-\Psi(P_{1,t-}^{k,l})
-\Psi'(P_{1,t-}^{k,l})\Gamma_1^{k,l}\Big]\lambda\nu(dy)dt\\
&=\Psi(0)+\E\int_0^T\Psi'(P_{1,t}^{k,l})\Big[2rP_{1,t}^{k,l}+F_1^{*,k}(t,P_1^k,\Lambda_{1}^k)+G_1^{*,k}(P_1^k,\Gamma_1^k,P_2,\Gamma_2)\\
&\qquad-F_1^{*,l}(t,P_1^l,\Lambda_{1}^l)-G_1^{*,l}(P_1^l,\Gamma_1^l,P_2,\Gamma_2)\Big]dt.
\end{align*}
Using the following facts:
\begin{align*}
&\Psi(0)=0, \ P_{1,t}^{k,l}\geq 0, \ \Psi'(P_{1,t}^{k,l})=e^{\kappa P_{1,t}^{k,l}}-1\geq 0, \\
&F_1^{*,k}\leq 0, \ G_1^{*,k}\leq 0, \ F_1^{*,l}\geq F_1^{*}, \ G_1^{*,l}\geq G_1^{*},
\end{align*}
we obtain
\begin{align*}
&\quad\E[\Psi(P_{1,0}^{k,l})]+\frac{1}{2}\E\int_0^T\Psi''(P_{1,t}^{k,l})|\Lambda_1^{k,l}|^2dt\\
&\qquad\qquad+\E\int_0^T\int_{\R_+}\Big[\Psi(P_{1,t-}^{k,l}+\Gamma_{1,t}^{k,l})-\Psi(P_{1,t-}^{k,l})
-\Psi'(P_{1,t-}^{k,l})\Gamma_1^{k,l}\Big]\lambda\nu(dy)dt\\
&\leq\E\int_0^T\Psi'(P_{1,t}^{k,l})\Big[2rP_{1,t}^{k,l}
-F_1^{*}(t,P_1^l,\Lambda_{1}^l)-G_1^{*}(P_1^l,\Gamma_1^l,P_2,\Gamma_2)\Big]dt.
\end{align*}
Keeping in mind $\vartheta\leq P^l_1, P^l_1+\Gamma^l_1\leq M$, $\vartheta\leq P_2, P_2+\Gamma_2\leq M$ are uniformly bounded, we have the following estimates for $F_1^{*}(t,P_1^l,\Lambda_{1}^l)$ and $G_1^{*}(P_1^l,\Gamma_1^l,P_2,\Gamma_2)$:
\begin{align*}
-F_1^{*}(t,P_1^l,\Lambda_{1}^l)&\leq -\inf_{v\in\R^m}\Big[P_1^l|\sigma^{\top}v|^2+2v^{\top}(P_1^l\mu+\sigma\Lambda_1^l)\Big]\\
&\leq \frac{1}{\vartheta}(P_1^l\mu+\sigma\Lambda_1^l)^{\top}(\sigma\sigma^{\top})^{-1}(P_1^l\mu+\sigma\Lambda_1^l)\\
&\leq c+c|\Lambda_1^l|^2\\
&\leq c+3c(|\Lambda_1^{k,l}|^2+|\Lambda_1^{k}-\Lambda_2|^2+|\Lambda_1|^2),
\end{align*}
and
\begin{align*}
-G_1^{*}(P_1^l,\Gamma_1^l,P_2,\Gamma_2)&\leq-\inf_{u\in\R}\Big[ \int_{\R_+}\Big[(P_1^l+\Gamma_1^l)\big[[(1-uy)^+]^2-1\big]+(P_2+\Gamma_2)[(1-uy)^-]^2\Big]\lambda \nu(dy)\\
&\qquad\qquad+2uP_1(b+\lambda b_Y)\Big]\\
&=-\inf_{u\in\R}\Big[ \int_{\R_+}\Big[(P_1^l+\Gamma_1^l)[(1-uy)^+]^2+(P_2+\Gamma_2)[(1-uy)^-]^2\Big]\lambda \nu(dy)\\
&\qquad\qquad+2uP_1^l(b+\lambda b_Y)\Big]+\int_{\R_+}(P_1^l+\Gamma_1^l)\lambda\nu(dy)\\
&\leq -\inf_{u\in\R}\Big[\vartheta\lambda(\sigma_Y^2u^2-2 u b_Y+1)+2u\vartheta(b+\lambda b_Y)\Big]+\lambda M\\
&\leq c,
\end{align*}
where $c>0$ is a large constant independent of $l$ and $\kappa$.
The above estimates lead to
\begin{align*}
&\quad\E[\Psi(P_{1,0}^{k,l})]+\E\int_0^T\Big(\frac{1}{2}\Psi''(P_{1,t}^{k,l})-3c\Psi'(P_{1,t}^{k,l})\Big) |\Lambda_1^{k,l}|^2dt\\
&\qquad\qquad+\E\int_0^T\int_{\R_+}\Big[\Psi(P_{1,t-}^{k,l}+\Gamma_{1,t}^{k,l})-\Psi(P_{1,t-}^{k,l})
-\Psi'(P_{1,t-}^{k,l})\Gamma_1^{k,l}\Big]\lambda\nu(dy)dt\\
&\leq\E\int_0^T\Psi'(P_{1,t}^{k,l})\Big[2rP_{1,t}^{k,l}
+2c+3c|\Lambda_1^{k}-\Lambda_2|^2+3c|\Lambda_1|^2)
\Big]dt.
\end{align*}

Take $\kappa=12c$. Then $\frac{1}{2}\Psi''(x)- 3c\Psi'(x)=3c(e^{\kappa x}+1)
=3c\Psi'(x)+6c>0$.
So the term
$$\sqrt{\frac{1}{2}\Psi''(P_{1,t}^{k,l})-3c\Psi'(P_{1,t}^{k,l})}$$
converges strongly to
$$\sqrt{3c\Psi' (P_{1,t}^{k}-P_{1,t})+6c},$$
as $l\rightarrow\infty$, and they are uniformly bounded. Therefore,
$$\sqrt{3c\Psi'(P_{1,t}^{k,l})+6c}\;\Lambda_1^{k,l}$$
converges weakly to
$$\sqrt{3c\Psi' (P_{1,t}^{k}-P_{1,t})+6c}\; (\Lambda_1^k-\Lambda_1).$$
By the mean value theorem and the boundedness of $P$ and $\Gamma$, we obtain
\begin{align}\label{AGamma}
\Psi(P_{1,t-}^{k,l}+\Gamma_{1,t}^{k,l})-\Psi(P_{1,t-}^{k,l})
-\Psi'(P_{1,t-}^{k,l})\Gamma_2^{k,l}&=\frac{1}{c}e^{cP_{1,t-}^{k,l}}
\Big[e^{c\Gamma_1^{k,l}}-c\Gamma_1^{k,l}-1\Big] \geq \ep|\Gamma_{1}^{k,l}|^2,
\end{align}
for some small constant $\ep>0$.
We then get from the last weak convergence and Fatou's lemma that
\begin{align*}
&\quad\E\int_0^T\big(3c\Psi' (P_{1,t}^{k}-P_{1})+6c\big)|\Lambda_1^{k}-\Lambda_1|^2ds\\
&\leq\varliminf_{l\rightarrow\infty}\E\int_0^T\big(3c\Psi' (P_{1,t}^{k,l})+6c\big)|\Lambda_1^{k,l}|^2ds\\
&\leq\E\int_0^T\Psi'(P_{1,t}^{k}-P_{1})\Big[2r(P_{1,t}^{k}-P_{1})
+2c+3c|\Lambda_1^{k}-\Lambda_1|^2+3c|\Lambda_1|^2)\Big]ds,
\end{align*}
that is,
\begin{align*}
&\quad\E\int_0^T6c|\Lambda_1^{k}-\Lambda_1|^2ds\\
&\leq\E\int_0^T\Psi'(P_{1,t}^{k}-P_{1})\Big[2r(P_{1,t}^{k}-P_{1})
+2c+3c|\Lambda_1|^2)
\Big]ds.
\end{align*}
By passing to the limit $k\rightarrow\infty$ and applying dominated convergence theorem, we have
\begin{align*}
\lim_{k\rightarrow\infty}\E\int_0^T|\Lambda_1^k-\Lambda_1|^2dt=0.
\end{align*}

Using \eqref{AGamma}, we can similarly get
\begin{align*}
\lim_{k\rightarrow\infty}\E\int_0^T\int_{\R_+}|\Gamma_1^k-\Gamma_1|^2\lambda\nu(dy)dt=0.
\end{align*}
Because the sequence $\Gamma_1^k$, $k=1,2,\cdots,$ is uniformly bounded in $\linnu $, we conclude that $\Gamma_1 \in \linnu $.

Along appropriate subsequence (which is still denoted by $(\Lambda_{1}^k, \Gamma_{1}^k)$) we may obtain a.e. a.s. convergence of
\begin{align*}
\int_t^T(\Lambda_{1}^k)^{\top}dW+\int_t^T\int_{\R_+}\Gamma_1^k(y)\tilde \gamma(ds,dy)\rightarrow \int_t^T (\Lambda_{1})^{\top}dW+\int_t^T\int_{\R_+}\Gamma_1(y)\tilde \gamma(ds,dy).
\end{align*}
We now turn to prove
\begin{align}\label{AFG}
\lim_{k\rightarrow\infty}\int_t^T\Big[2rP_1^k&+F_1^{*,k}(s,P_1^k,\Lambda_{1}^k)+G_1^{*,k}(P_1^k,\Gamma_1^k,P_2,\Gamma_2)\Big]ds\nn\\
&=\int_t^T \Big[2rP_1+F_1^*(s,P_1,\Lambda_{1})+G_1^*(P_1,\Gamma_1,P_2,\Gamma_2)\Big]ds.
\end{align}
Apparently,
\begin{align*}
|F_1^{*,k}(s,P_1^k,\Lambda_{1}^k)-F_1^*(s,P_1,\Lambda_{1})|&\leq |F_1^{*,k}(s,P_1^k,\Lambda_{1}^k)-F_1^*(s,P_1^k,\Lambda_{1}^k)|\\
&\qquad +|F_1^{*}(s,P_1^k,\Lambda_{1}^k)-F_1^*(s,P_1,\Lambda_{1})|.
\end{align*}
Recall that $\Lambda_{1,s}^k\rightarrow\Lambda_{1,s}$ a.e., a.s., so there exists $k_{1}(\omega,s)$ such that $|\Lambda_{1,s}^k|\leq 1 +|\Lambda_{1,s}|$ for $k\geq k_{1}$.
Notice that
\begin{align*}
F_1(s,v,P_1^k,\Lambda_{1}^k)\geq \vartheta\delta|v|^2-2c_1|v||\Lambda_1^k|
\geq \vartheta\delta|v|^2-2c_1|v|(1+|\Lambda_1|)>0,
\end{align*}
if $|v|>c_2(1 +|\Lambda_{1,s}|)$ with $c_2>0$ being sufficiently large. Hence, for $k\geq\max\{ c_2(1+|\Lambda_{1,s}|), k_{1}\}$, we have
\begin{align*}
F_1^*(s,P_1^k,\Lambda_{1}^k)
&=\inf_{\substack{v\in\Pi\\|v|\leq c_2(1 +|\Lambda_{1,s}|)}}F_1(s,v,P_1^k,\Lambda_{1}^k)
\geq \inf_{\substack{v\in\Pi\\|v|\leq k}}F_1(s,v,P_1^k,\Lambda_{1}^k)
=F_1^{*,k}(s,P_1^k,\Lambda_{1}^k).
\end{align*}
We also have the reverse inequality $F_1^*\leq F_1^{*,k}$ by definition.
Therefore,
\begin{align*}
\lim_{k\rightarrow\infty}|F_1^{*,k}(s,P_1^k,\Lambda_{1}^k)-F_1^*(s,P_1^k,\Lambda_{1}^k)|=0.
\end{align*}
By its concavity, hence continuity of $F_1^*$ w.r.t. $(P_1,\Lambda_1)$ on $[\frac{\vartheta}{2},2M]\times\R^m$, we have
$$\lim_{k\rightarrow\infty}|F_1^{*}(s,P_1^k,\Lambda_{1}^k)-F_1^*(s,P_1,\Lambda_{1})|=0.$$

In the same manner we can prove that
\begin{align*}
\lim_{k\rightarrow\infty}|G_1^{*,k}(s,P_1^k,\Gamma_{1}^k,P_2,\Gamma_2)-G_1^*(P_1,\Gamma_{1},P_2,\Gamma_2)|=0.
\end{align*}
Since
\begin{align*}
|F_1^{*,k}(t,P_1^k,\Lambda_{1}^k)|\leq c(1+|\Lambda_{1}^k|^2), \ |G_1^{*,k}(P_1^k,\Gamma_{1}^k,P_2,\Gamma_2)|\leq c,
\end{align*}
the dominated convergence theorem leads to \eqref{AFG}.

Now it is standard to show that
$$\lim_{k\rightarrow\infty}\E[\sup_{t\in[0,T]}|P^k_{1,t}-P_{1,t}|]=0.$$
Please refer to Antonelli and Mancini \cite {AM} for details.

\end{document}